\documentclass[aps,prb,twocolumn,amsmath,amssymb,showpacs,superscriptaddress]{revtex4}
\usepackage{graphicx}
\usepackage{amssymb}
\usepackage{natbib}
\usepackage{calc}
\usepackage{SIunits}
\usepackage{textcomp}
\usepackage{color}

\begin{document}

\title{Spin Fluctuations in the Rare-Earth Doped Bilayer Nickelates}

\author{Honglin Zhou}
\thanks{These authors made equal contributions to this paper}
\affiliation{Beijing National Laboratory for Condensed Matter Physics, Institute of Physics, Chinese Academy of Sciences, Beijing 100190, China}
\affiliation{School of Physical Sciences, University of Chinese Academy of Sciences, Beijing 100190, China}
\author{Xinman Ye}
\thanks{These authors made equal contributions to this paper}
\affiliation{Institute of Neutron Science and Technology, Guangdong Provincial Key Laboratory of Magnetoelectric Physics and Devices,
School of Physics, Sun Yat-Sen University, Guangzhou 510275, China}
\author{Gang Wang}
\thanks{These authors made equal contributions to this paper}
\affiliation{Beijing National Laboratory for Condensed Matter Physics, Institute of Physics, Chinese Academy of Sciences, Beijing 100190, China}
\affiliation{School of Physical Sciences, University of Chinese Academy of Sciences, Beijing 100190, China}
\author{Devashibhai Adroja}
\thanks{These authors made equal contributions to this paper}
\affiliation{ISIS Neutron and Muon Source, Rutherford Appleton Laboratory, Chilton, Didcot OX11 0QX, United Kingdom}
\affiliation{Physics Department, University of Johannesburg, Auckland Park 2006, South Africa}
\author{David Tam}
\affiliation{Institut Laue-Langevin, 71 Avenue des Martyrs, 38000 Grenoble, France}
\author{Michael Marek Koza}
\affiliation{Institut Laue-Langevin, 71 Avenue des Martyrs, 38000 Grenoble, France}
\author{Zhilun Lu}
\affiliation{School of Chemical and Process Engineering, University of Leeds, Leeds, West Yorkshire LS2 9JT, UK}
\author{Jinguang Cheng}
\thanks{jgcheng@iphy.ac.cn}
\affiliation{Beijing National Laboratory for Condensed Matter Physics, Institute of Physics, Chinese Academy of Sciences, Beijing 100190, China}
\affiliation{School of Physical Sciences, University of Chinese Academy of Sciences, Beijing 100190, China}
\author{Dao-Xin Yao}
\thanks{yaodaox@mail.sysu.edu.cn}
\affiliation{Institute of Neutron Science and Technology, Guangdong Provincial Key Laboratory of Magnetoelectric Physics and Devices,
School of Physics, Sun Yat-Sen University, Guangzhou 510275, China}
\author{Huiqian Luo}
\thanks{hqluo@iphy.ac.cn}
\affiliation{Beijing National Laboratory for Condensed Matter Physics, Institute of Physics, Chinese Academy of Sciences, Beijing 100190, China}

\date{\today}
\pacs{74.70.-b, 75.40.Gb, 75.30.Et, 74.25.Ha, 78.70.Nx}

\begin{abstract}
Spin fluctuations have been generally believed as the pairing glue of high-$T_c$ superconductivity. Recent inelastic neutron scattering (INS) studies have revealed a weak flat spin-fluctuation signal around 45 meV in the bilayer nickelate La$_3$Ni$_2$O$_{7-\delta}$, suggesting strong interlayer and weak intralayer magnetic couplings ($SJ_{\perp}\approx$ 60 meV, $SJ_{\parallel}\leq$ 3.5 meV) in contrast to cuprate and pnictide superconductors. Here, we report further INS studies on the Pr and Nd doped La$_3$Ni$_2$O$_{7-\delta}$ powder samples at ambient pressure. Besides the crystalline electric field excitations at low energies, we have found that the 45 meV flat mode splits into two modes in doped compounds, along with another weak mode at about 60 meV, where the spin fluctuations in La$_2$NdNi$_2$O$_{7-\delta}$ are stronger than La$_3$Ni$_2$O$_{7-\delta}$ and La$_2$PrNi$_2$O$_{7-\delta}$. Our results are consistent with an enhanced interlayer coupling $SJ_{\perp}$ within the stripe-type Heisenberg model framework, where the estimated $SJ_{\perp}$ value is in the range of about 69 to 73 meV for the rare-earth doped bilayer nickelates.
\end{abstract}

\keywords{nickelate superconductors, high temperature superconductors, spin fluctuations, exchange couplings, inelastic neutron scattering}

\maketitle

\section{Introduction}
Spin fluctuations are arguably the common thread in understanding the pairing mechanism of high-$T_c$ superconductivity\cite{djscalapino2012,jttranquada2014,bdwhite2015,pdai2012,pdai2015,hdcjohnston2010,xhchen2014,achristianson2008,qgu2022,ywang2025,zwang2025}. In cuprates, while the spin waves emerge at the N\'{e}el-type antiferromagnetic (AF) wavevector ($\mathbf{Q}_{\mathrm{AF}}$) in the parent compound and persist to high energy over than 200 meV\cite{jttranquada2014,pdai2012}, the low-energy spin fluctuations form a collective spin resonance mode in the superconducting state of doped compounds\cite{meschrig2006}. The intensity of resonance mode behaves like a superconducting order parameter with a peak energy ($E_R$) linearly scaling with $T_c$ and a downward dispersion confined by the $d-$wave superconducting gap\cite{ysidis2007}. Similar scenario is established in the iron-based superconductors, where both strong spin fluctuations and spin resonance have been extensively discovered around the stripe-type wavevector of iron pnictides\cite{pdai2012,pdai2015,hdcjohnston2010,xhchen2014,achristianson2008}, and the resonance dispersion turns to depend on the fermiology in the multi-band $s\pm$-wave pairing picture\cite{rzhang2018,txie2021,wshong2020,wshong2023,yli2025,zzli2025}. Notably, the spin wave dispersion in the AF ordered parent compounds can be described by an effective Heisenberg model with dominated in-plane exchange couplings ($J_{1a},J_{1b},J_{2}$) and a weak out-of-plane coupling ($J_c$)\cite{pdai2012,pdai2015}. The spin fluctuations keep such memory in the superconducting compounds but become quasi-two-dimensional (quasi-2D) like with nearly zero $J_c$\cite{pdai2015,hdcjohnston2010,xhchen2014}. Both the high energy scale determined by the in-plane exchange couplings of local moments and the strong coupling with itinerant electrons in the spin fluctuations are essential to understand the high-$T_c$ mechanism\cite{zzli2025,fwang2011,mwang2013}.

The bilayer nickelates La$_{3-x}$$R_x$Ni$_2$O$_{7-\delta}$ ($R$= Pr, Nd, Sm,...) recently catalogued as a new family of high-$T_c$ superconductors\cite{hsun2023,yzhang2023,nnwang2024,jhou2023,gwang2024,ekko2024,gzhou2025,wwu2024,ychen2025}, provide a distinct vision of high-$T_c$ mechanism related to spin fluctuations\cite{mwang2024,zliu2024,zliao2023,yyang2023,qqin2023,qqin2024,xzqu2024,jyang2024,kaneko2024,ouyang2024,ryee2024,kjiang2024,clu2024,jli2025,yhcao2026}.
The La$_3$Ni$_2$O$_{7}$ compound, referred as the bilayer Ruddlesden-Popper phase, crystallizes in an orthorhombic $Amam$ structure with alternating stacks of two NiO$_2$ planes and tilted octahedra\cite{mwang2024}. The nominal valence state is Ni$^{2.5+}$ for $3d^{7.5}$ electronic configuration with half-filled $d_{z^2}$ and quarter-filled $d_{x^2-y^2}$ orbitals near the Fermi level\cite{jyang2024,kaneko2024,ouyang2024,ryee2024,kjiang2024,clu2024,jli2025,yhcao2026}. With increasing pressure, the bond angle of Ni-O-Ni along $c$ axis changes from 168$^{\degree}$ to 180$^{\degree}$, leading to an orthorhombic $Fmmm$ or tetragonal $I4/mmm$ phase showing a high-$T_c$ superconductivity up to 80 K\cite{hsun2023,ychen2025,jli2025}. Although the microscopic mechanism remains inconclusive, the interlayer coupling is generally believed as a crucial key responsible for the superconducting pairing\cite{ryee2024,kjiang2024,clu2024}. Possible magnetic orders are also proposed by neutron diffraction, positive muon spin relaxation ($\mu^{+}$SR), nuclear magnetic resonance (NMR) and Raman scattering measurements\cite{taniguchi1995,ykobayashi1996,cdling2000,gwu2001,liu2023,iplokhikh2025,khasanov2024,kchen2024,kchen2025,ni2025,dzhao2024,jluo2025,kfan2026,ghe2026}. Indeed, the physical pressure probably drives the emergence of a hole pocket of $d_{z^2}$ orbital, or uniformly enhances the bandwidth, in either case it will yield a predominant $s\pm$-wave pairing triggered by spin fluctuations\cite{lechermann2023,yzhang2023a,yzhang2023b,hoh2023,luo2024,sakakibara2024,ybliu2023,qgyang2023,tian2024,heier2024,yhgu2025,jhji2024,jyyou2025,jzhan2025,kyjiang2025}. The rare earth doping provides additional chemical pressure to modify the interlayer coupling and the $d_{z^2}$ orbital electronic structure\cite{pan2024,mzhang2024,gwang2025,huo2025,jgyang2024,yzhang2023c,zdong2025,cqchen2025,gdzhou2026}, which could potentially enhance $T_c$ as demonstrated in La$_2$SmNi$_2$O$_{7-\delta}$ and La$_{0.6}$Nd$_{2.4}$Ni$_2$O$_{7-\delta}$\cite{fyli2024a,fyli2025b,qzhong2025,jjfeng2024,zzqiu2025}.

\begin{figure}[tbp]
\includegraphics[width=0.47\textwidth]{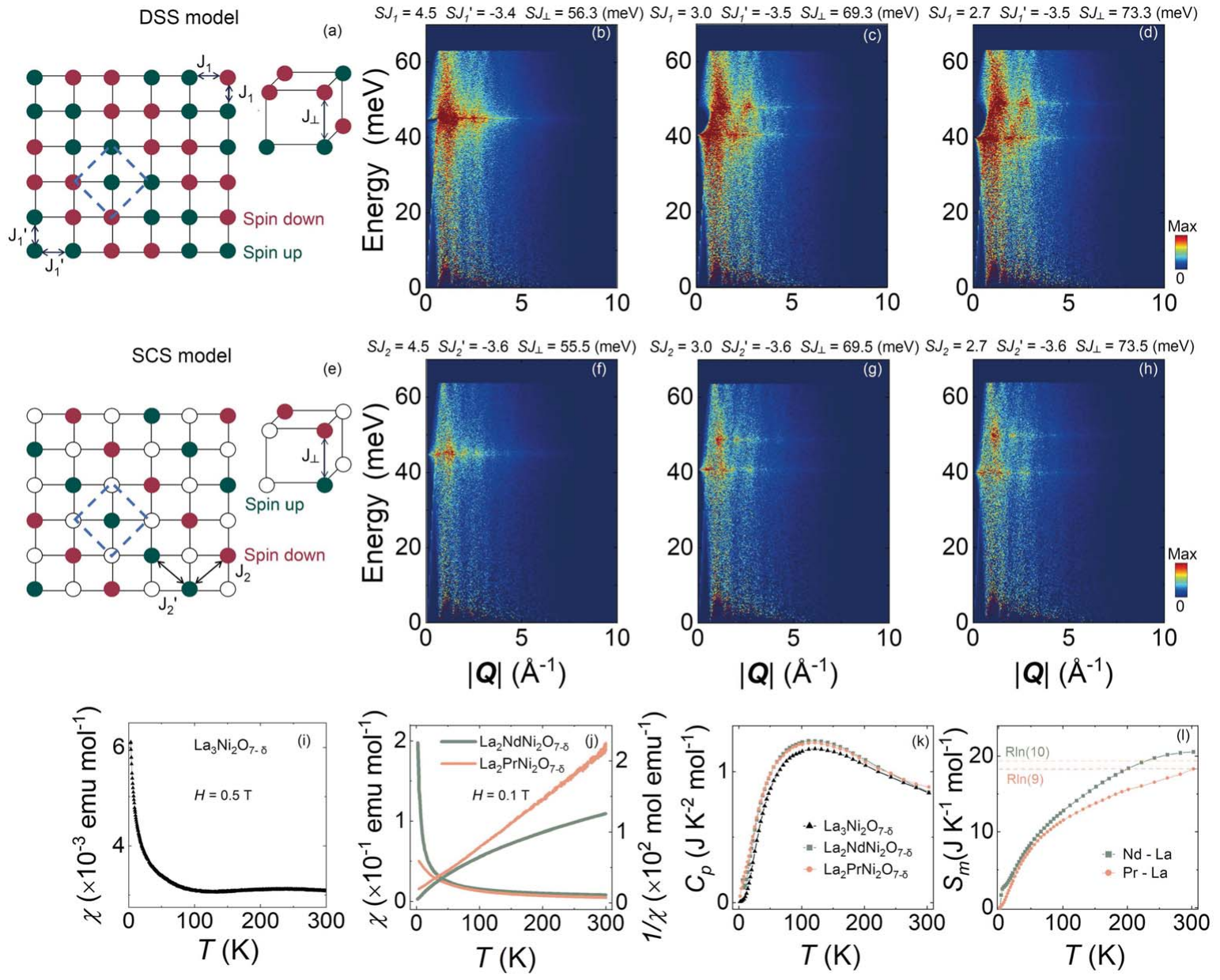}
\caption{
(a)-(h) SpinW calculations based on the double spin stripe (DSS) and the single spin-charge stripe (SCS) AF orders using different combinations of exchange couplings. The nearest-neighbor intralayer exchange couplings $J_1$,$J_1^{\prime}$ in DSS model (or $J_2$,$J_2^{\prime}$ in SCS model) and interlayer exchange coupling $J_{\perp}$ are indicated in (a) and (e). The green, red and white balls represent the spin up, spin down and nonmagnetic Ni atoms, respectively\cite{xchen2024}.
(i) Magnetic susceptibility $\chi$ of La$_3$Ni$_2$O$_{7-\delta}$. 
(j) $\chi$ and its inverse $1/\chi$ of La$_2$NdNi$_2$O$_{7-\delta}$ and La$_2$PrNi$_2$O$_{7-\delta}$.
(k) Heat capacity $C_p$ versus $T$ for three compounds.
(l) Magnetic entropy of Pr and Nd doped samples from the integration of $C_m/T$. Here $C_m$ is obtained by subtracting the $C_p$ of La$_3$Ni$_2$O$_{7-\delta}$.
}
\end{figure}

Here we report an inelastic neutron scattering (INS) study on the  Nd and Pr doped bilayer nickelates La$_2$PrNi$_2$O$_{7-\delta}$ and La$_2$NdNi$_2$O$_{7-\delta}$ powder samples at ambient pressure.
Previous INS on La$_3$Ni$_2$O$_{7-\delta}$ powder sample has revealed a weak flat spin-fluctuation signal around 45 meV, which could be interpreted as a result of strong interlayer ($SJ_{\perp}\approx$ 60 meV) and weak intralayer ($SJ_{\parallel}\leq$ 3.5 meV) magnetic couplings for stripe-type AF orders\cite{txie2024}. Such conclusion is confirmed by the measurements of resonant inelastic X-ray scattering (RIXS) and INS on single-crystalline samples\cite{xchen2024,jzhao2025}. In Pr and Nd doped samples, we have found that the $SJ_{\perp}$ is further enhanced to about 70 meV. The main feature is two splitting flat modes of spin fluctuations around 45 meV, which can be theoretically simulated based on the stripe-type AF orders by only considering the nearest-neighbor exchange couplings (Fig. 1(a)-(h)). Such enhancement of $J_{\perp}$ possibly accounts for the increased $T_c$ from 80 K to near 100 K in rare-earth doped La$_3$Ni$_2$O$_{7-\delta}$ \cite{fyli2025b,zzqiu2025}.

\section{Results and discussion}
\subsection{Sample characterization}
Powder sample of La$_2$LnNi$_2$O$_{7-\delta}$ (Ln=La, Pr, Nd) are prepared by the sol-gel method\cite{gwang2024,gwang2025,zhang1994}, the sample purity is presented in Supplemental Materials. The magnetic susceptibility $\chi$ and heat capacity $C_p$ results are presented in Fig. 1 (i)-(l). No clear anomaly of $\chi$ is detected for pure La sample, consistent with previous reports on powder samples\cite{taniguchi1995,ykobayashi1996,cdling2000,gwu2001,liu2023,iplokhikh2025,khasanov2024,kchen2024}. The $\chi$ of Nd samples show a Curie-Weiss (CW) behavior as indicating by the linear $1/\chi$ vs $T$ (Fig. 1(i)(j)), CW-fitting of $\chi$ below 50 K gives the effective magnetic moments $\mu_{\mathrm{eff}}$(Nd)=2.8 $\mu_\mathrm{B}$ and $\mu_{\mathrm{eff}}$(Pr)=3.7 $\mu_\mathrm{B}$, consistent with the $4f$ electron contributions. No clear phase transition can be identified in $C_p$ except a very broad hump around 100 K for all three samples, thus any potential magnetic ordering of Ni is likely too weak to produce a detectable anomaly in either thermodynamic or magnetic measurements. Overall, all three samples have similar temperature dependence of $C_p$, suggesting almost the same contribution from phonons. Thus the magnetic heat capacity $C_m$ (or 4f-contribution $C_{4f}$) can be deduced by deducting $C_p$ of La sample, and the magnetic entropy $S_m$ can be obtained by integrating $C_m/T$. The crystalline electric field (CEF) contribution from Pr$^{3+}$ ($4f^2, J=4$) and Nd$^{3+}$ ($4f^3, J=9/2$) atoms is expected to yield a plateau at $R\ln(9)$=17.7 J mol$^{-1}$K$^{-1}$ and $R\ln(10)$=19.1 J mol$^{-1}$K$^{-1}$, respectively\cite{lyang2025}. However, this is not the case in Fig. 1(l), where $S_m$(Nd-La) exceeds the plateau value, suggesting additional contribution from spin fluctuations. It should be noticed that the absence of anomaly in $\chi$ and $C_p$ of Pr and Nd doped samples cannot rule out the possibility of a weak magnetic order of Ni. In fact, $\mu^{+}$SR measurements have estimated $m_{\mathrm{Ni}}\approx$ 0.22/0.42 $\mu_\mathrm{B}$ when the magnetic moments align along $c/ab$ axis for pure La sample\cite{kchen2024}, and the neutron powder diffraction give $m_{\mathrm{Ni}}\approx$ 0.15/0.85 $\mu_\mathrm{B}$ on low and high-moment sites of La$_2$PrNi$_2$O$_{7}$\cite{iplokhikh2025}. However, it is clear that the rare-earth ions do not form any long-range magnetic order at low temperature in Ruddlesden-Popper nickelates.

\subsection{Inelastic neutron scattering}
To search the spin excitations of La$_2$NdNi$_2$O$_{7-\delta}$, we have performed INS measurements at the time-of-flight spectrometer MERLIN, ISIS Neutron and Muon Source of UK\cite{bewley2006,isisExp2025}. We summarize the raw data in Fig. 2(a)-(h) by plotting the 2D slice of energy versus the absolute momentum transfer $\mid\mathbf{Q}\mid$ (referred as $Q$ in the following) for different incident neutron energies $E_i=$ 15, 24, 50 and 79 meV measured at $T=5$ and 110 K. Nearly $Q$-indenpendent CEF excitations are observed at the energy transfer $E=$ 5.5 and 22 meV, which feature as peaks in 1D $E$-cuts in the 5 K- 100 K data sets (Fig. 2(i)-(k)) with stronger intensity at 5 K. Although the data set of $E_i=$ 79 meV is contaminated by strong phonon scattering from the sample and aluminum can at high $Q$ region, the difference between 5 K and 110 K 1D $E$-cut shows three new peak features at $E=$ 43, 48 and 60 meV,  which are possible spin excitations similar to undoped compound (Fig. 2(l))\cite{txie2024,xchen2024,jzhao2025}. To confirm the magnetic origin of these features, we then integrate the intensity at 5 K focusing on these specific energy windows and plot the 1D $Q$-cuts in Fig. 2(m)-(p). Most of them show a decreasing intensity by increasing $Q$, a typical evidence for magnetic scattering. The abrupt increase of the 1D cut $E=$[21, 23] meV in Fig. 2(o) is due to phonon backgrounds at high $Q$. The positions of Bragg peaks can be found in the quasi-elastic cut with $E=$[-0.5, 0.5] meV for $E_i$= 15 meV measurement (Fig. 2(m)). No clear magnetic peaks could be identified, and most nuclear peaks locate above 2 \AA$^{-1}$. Thus the broad peak of the intensity with $E=$[1.5, 2.5] meV from $Q$= 0.5 to 2 \AA$^{-1}$ could be also spin excitations emerging from low energy, since only two tiny nuclear peaks exist in this small $Q$ region\cite{txie2024}. No signals related to any spin excitations above 70 meV are observed in the $E_i=$ 160 meV data, which is consistent with the RIXS and INS results on single crystals\cite{xchen2024,jzhao2025}.

\begin{figure}[t]
\includegraphics[width=0.47\textwidth]{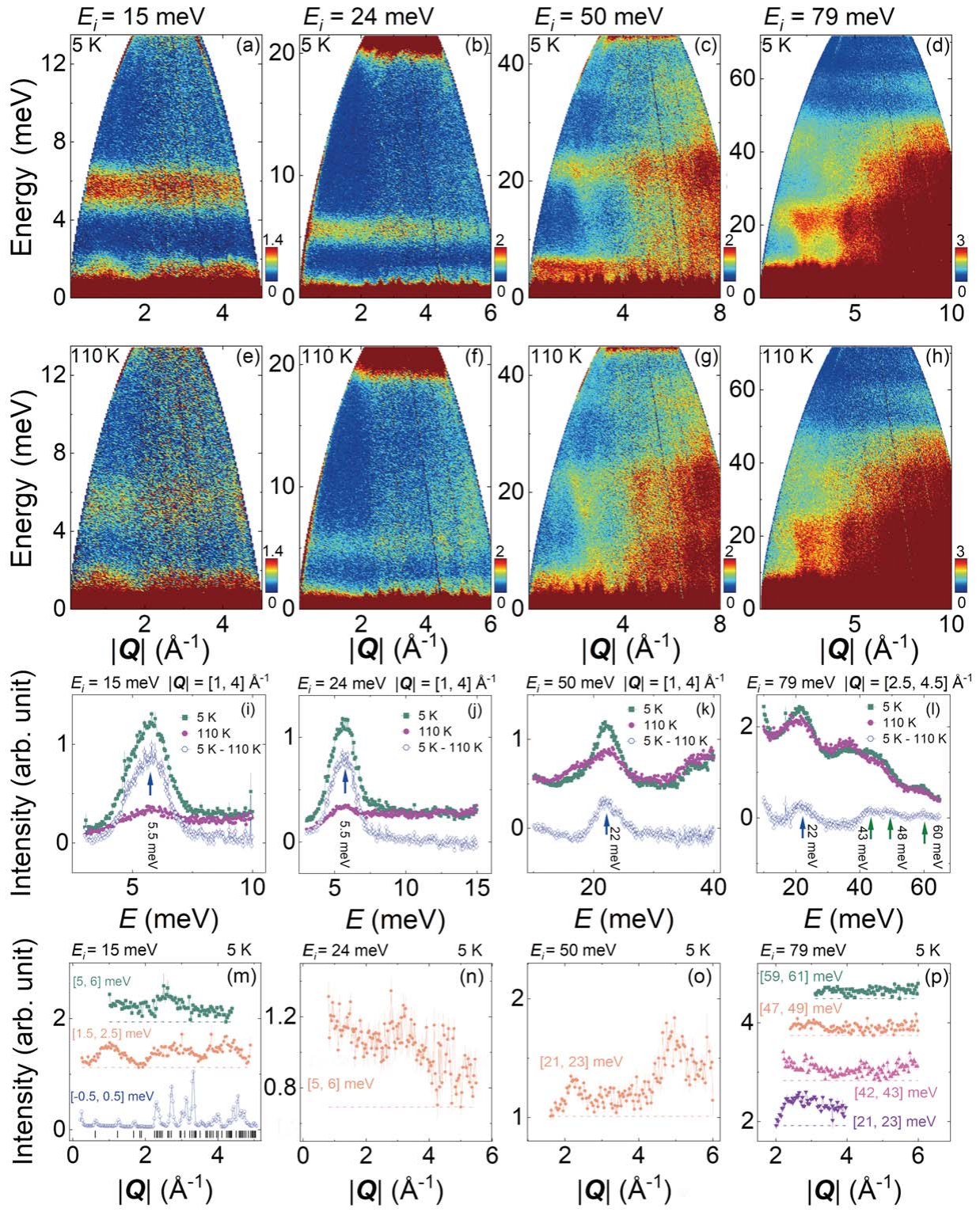}
\caption{
INS spectra of La$_2$NdNi$_2$O$_{7-\delta}$ collected at MERLIN.
(a)-(d) Data measured at 5 K with incident energy $E_i=$ 15, 24, 50 and 79 meV.
(e)-(h) Data measured at 110 K with similar $E_i$s.
(i)-(l) $E$-cuts at $T=$ 5 K, 110 K, and their differences with Bose corrections(See Supplemental Materials). The blue arrows mark the CEF excitations, and the green arrows mark the possible spin excitations.
(m)-(p) $Q$-cuts at $T=$ 5 K at the typical energy windows of excitations. The data is vertically shifted for clarity. The elastic cut with $E=$[-0.5, 0.5] meV is also presented in (m), where all nuclear peaks are marked by vertical bars.
}
 \end{figure}

\begin{figure}[t] \centering
\includegraphics[width=0.47\textwidth]{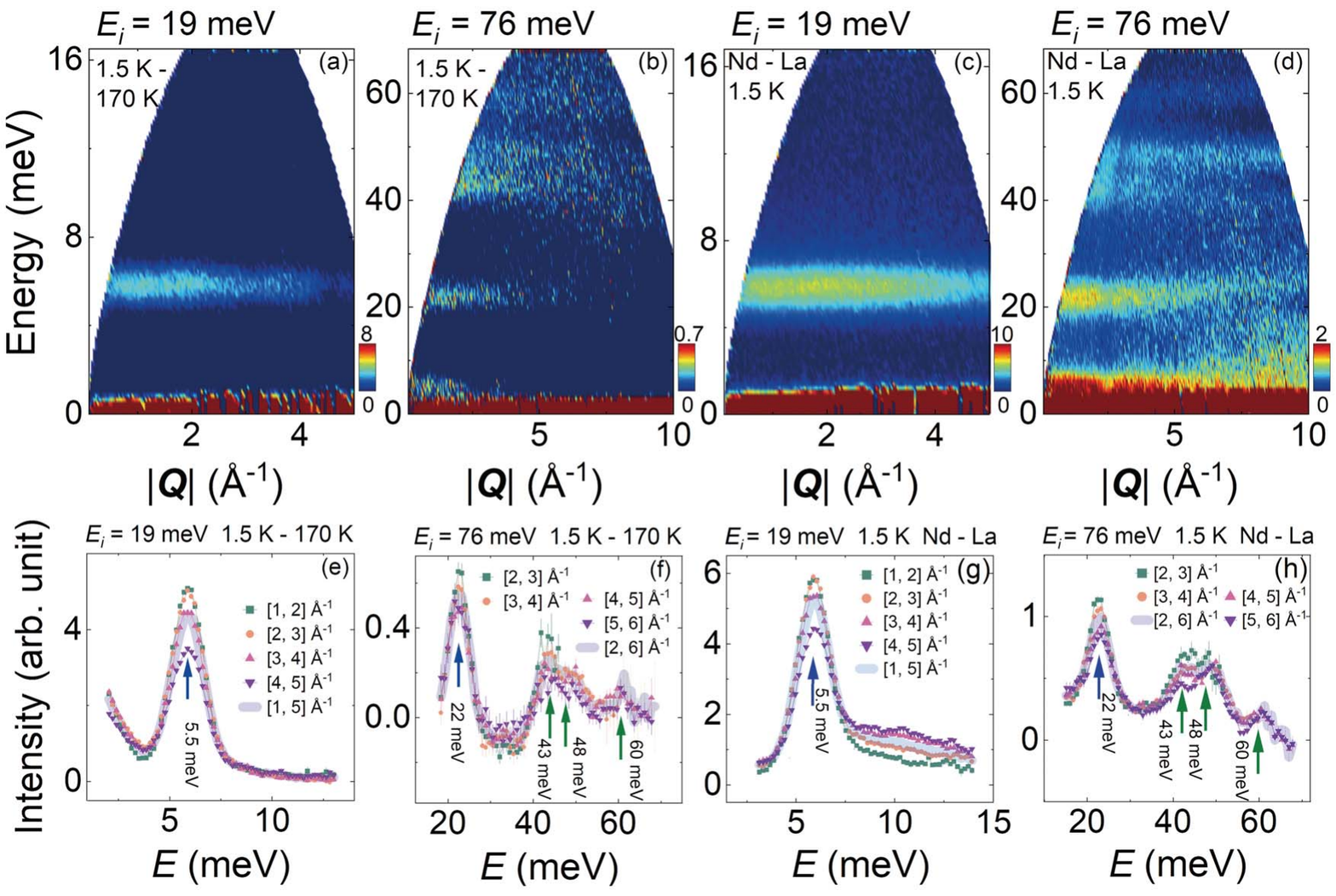}
\caption{
INS spectra of La$_2$NdNi$_2$O$_{7-\delta}$ collected at PANTHER with $E_i=$ 19 and 76 meV.
(a) and (b) Spectra after subtracting 170 K from 1.5 K data. 
(c) and (d) Spectra after subtracting La$_3$Ni$_2$O$_{7-\delta}$ from La$_2$NdNi$_2$O$_{7-\delta}$ data at $T=1.5$ K.
(e)-(h) $E$-cuts at different $Q$ regions corresponding to (a)-(d). Bose corrections are applied to (e) and (f)(See Supplemental Materials). The blue arrows mark the CEF excitations at 5.5 and 22 meV, and the green arrows mark the spin excitations at 43, 48 and 60 meV.
}
\end{figure}

Further INS measurements on three samples La$_2$LnNi$_2$O$_{7-\delta}$ (Ln=La, Pr, Nd) were carried out at the time-of-flight spectrometer PANTHER with incident neutron energy $E_i=$ 12.5, 19 and 76 meV (See Supplemental Materials)\cite{Panther2022,illExp2025}. Fig. 3 presents the results of La$_2$NdNi$_2$O$_{7-\delta}$ collected at PANTHER. Again, the energy levels of CEF excitations can be easily observed at $E=$ 5.5 and 22 meV from the 1D $E-$cut by subtracting the 170 K data from 1.5 K data (Fig. 3(a), (b), (e) and (f)). The original peak positions are actually at 6 meV and 23 meV at $T=1.5$ K with stronger intensities than 170 K data, which can be roughly fitting by a CEF model of Nd$^{3+}$. The high $Q$-dependence of the 6 meV mode basically follows the square of magnetic form factor $f(Q)^2$, but the 23 meV mode is mixed by phonon excitations (See Supplemental Materials). However, the splitting signals at $E=$ 43 and 48 meV cannot fully described by the Nd$^{3+}$ CEF model. As they are very similar to the case of flat 45 meV mode in La$_3$Ni$_2$O$_{7-\delta}$, it is reasonable to believe that they are magnon excitations. Another feature at $E=$ 60 meV is also observed (Fig. 3(b) and (f)). To double check the magnetic nature of these signals, we then take the data set of La$_3$Ni$_2$O$_{7-\delta}$ measured at the same condition as the reference of phonon background, since the 45 meV signal in the undoped sample is much weaker than the Nd-doped sample, but the phonon contribution at high Q is similar. The subtraction of Nd - La data set gives promising results for all above mentioned energy levels, especially for two splitting modes around 45 meV. The estimated energy resolution of $E_i=$ 76 meV around 45 meV is within 2.3 meV, much smaller than the splitting energy (5 meV). Thus we confirm unambiguously that the splitting phenomenon in the excitation spectrum is an intrinsic characteristic of this system. It seems that the splitting of spin excitations around 45 meV develops upon increasing $Q$, which is an unique feature of the spin excitations in bilayer nickelates\cite{txie2024}, and has never been observed in cuprate or pnictide superconductors\cite{djscalapino2012,jttranquada2014,bdwhite2015,pdai2012,pdai2015}.

After establishing the excitations in La$_2$NdNi$_2$O$_{7-\delta}$, we then present the INS results of La$_2$PrNi$_2$O$_{7-\delta}$ by processing similar analyses (Fig. 4). Given that Pr$^{3+}$ is a non-Kramers ion with
multiplet state\cite{lyang2025}, the CEF excitations are populated in 4 doublet levels with broad distributions at high temperature. Possible energy levels could be identified at $E=$ 2, 3.5, 4.2 and 6 meV from the 1.5 K and subtracted data (Fig. 4(a),(b),(e) and (f)), and they can be roughly described by the CEF model of Pr$^{3+}$. No magnetic peaks could be identified in the elastic $Q$-cut, and the CEF excitations decrease with increasing $Q$ and follow the square of magnetic form factor $f(Q)^2$ of Pr$^{3+}$ at high $Q$ range (Fig. 4(i) and (j)). The splitting of spin excitations around 45 meV still exists, but becomes weak at about $E=$ 44 and 47 meV, along with weaker signals at $E=$ 60 meV (Fig. 4(c), (d), (g) and (h)). It is hard to distinguish the $Q$-dependence of magnetic and phonon scatterings by doing similar subtraction (Fig. 4(k) and (l)) (See Supplemental Materials). Further measurements on single crystals are highly desired.

\begin{figure}[t] \centering
\includegraphics[width=0.47\textwidth]{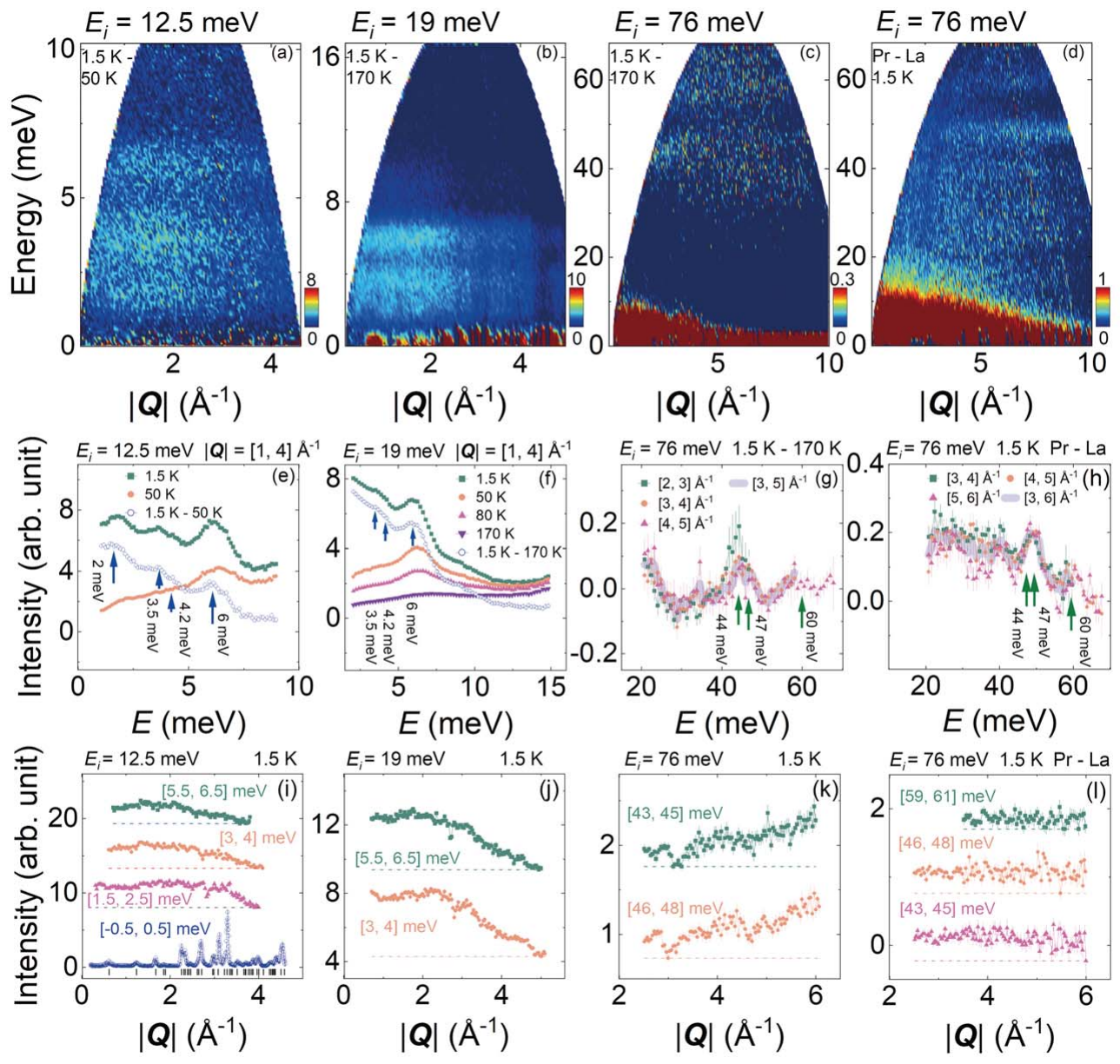}
\caption{
INS spectra of La$_2$PrNi$_2$O$_{7-\delta}$ collected at PANTHER with $E_i=$ 12.5, 19 and 76 meV.
(a) Spectrum of subtracting 50 K from 1.5 K data ($E_i=$ 12.5 meV).
(b)(c) Spectra after subtracting 170 K from 1.5 K data ($E_i=$ 19 and 76 meV).
(d) Spectra for $E_i=$ 76 meV after subtracting La$_3$Ni$_2$O$_{7-\delta}$ from La$_2$PrNi$_2$O$_{7-\delta}$ data at $T=1.5$ K.
(e)-(h) $E$-cuts at different regions corresponding to (a)-(d). Bose corrections are applied to panel (e), (f) and (g)(See Supplemental Materials). The blue arrows mark the CEF excitations at 2, 3.5 and 6 meV, and the green arrows mark the spin excitations at 44 meV, 47 meV and 60 meV.
(i)-(l) $Q$-cuts at $T=$ 1.5 K by focusing energy windows. The elastic cut with $E=$[-0.5, 0.5] is also presented in panel (i), where all nuclear peaks are marked by vertical bars.
}
\end{figure}

\subsection{Theoretical analyses}
Finally, we discuss the INS results using four theoretical models: single spin-charge stripe order (SCS) , double spin stripe order (DSS) , A-type AF order (AFM-A), and G-type AF order (AFM-G). We have performed numerical calculations of spin excitation spectra based on the linear spin wave theory and the SpinW software package (See Supplemental Materials). According to previous reports, the stripe-type AF orders SCS and DSS are the most likely cases in the bilayer nickelates: where Ni$^{2+}$ spin and Ni$^{3+}$ charge stripes diagonally intertwine in SCS (Fig. 1(e)), or doubly ordered without charge inhomogeneity in DSS (Fig. 1(a)), both of them have the same wavevector $\mathbf{Q}_{\mathrm{AF}}=$ (0.25, 0.25) in the quasi-tetragonal lattice\cite{iplokhikh2025,xchen2024}. Therefore, by considering the nearest-neighbor interlayer and intralayer interactions, we can construct a simple Heisenberg Hamiltonian \cite{tkaneko2025,hoh2026,yzhong2026}:
\begin{equation}
	H=J_{\alpha}\sum_{<i,j>} {S_{i}}\cdot {S_{j}}+J_{\alpha}'\sum_{<i,j>'} {S_{i}}\cdot {S_{j}}+J_{\bot }\sum_{i} {S^{t}_{i}}\cdot {S^{b}_{i}},
\end{equation}
where $J_{\alpha}$ and $J_{\alpha}^{\prime}$ are the AF and ferromagnetic (FM) intralayer nearest-neighbor interactions, respectively. We use $\alpha=1, 2$ to denote the case of DSS and SCS models for their different bonding lengths (3.8 \AA\ and 5.4 \AA). $J_\perp$ is the interlayer nearest-neighbor interaction at the same site with top and bottom layer spins ($S^{t}$, $S^{b}$), where the effective spin $S$ arises from the total spin contributions of different orbitals. It should be noticed that all the exchange parameters are effective $J_{eff}$ calculated from the eigen energy of a full Hamiltonian, where the joint effects of CEF effect from Ni$^{n+}$, orbital hopping and Coulomb repulsion are considered. Since the in-plane lattice is not exactly tetragonal, it is reasonable to have different absolute value of $J_{\alpha}$ and $J_{\alpha}^{\prime}$. For the stripe-type models, we adopted the parameters shown in Table I to describe the INS data of all three samples, the obtained spectra are presented in Fig.1(b)-(d) and (f)-(h).
Based on the simulation results of the DSS and SCS models, it can be clearly observed that the interlayer interaction ($SJ_\perp$) is much stronger than the intralayer interaction ($SJ_{\alpha}$, $SJ_{\alpha}^{\prime}$). In the doped case, $SJ_\perp$ increases significantly, while the intralayer interaction undergo changes, resulting in a band gap between the acoustic and optical branches in DSS model, or anisotropic in-plane bandtop in SCS model. It finally turns to the occurrence of splitting flat modes around 45 meV in our powder INS measurements. Such phenomenon is more pronounced in La$_2$NdNi$_2$O$_{7-\delta}$ due to its larger $J_\perp$. On the other hand, the AFM-A and AFM-G models with isotropic intralayer interactions cannot produce the splitting flat-band like spectrum but only wave-like characteristics, thus they can be ruled out in the bilayer nickelate system (See Supplemental Materials). 
\begin{table}[!ht]
	\centering
	\caption{Model parameters for DSS (top) and SCS (bottom)}
	\begin{tabular}{|c|c|c|c|}
		\hline
		Couplings &La$_3$Ni$_2$O$_{7-\delta}$ &La$_2$PrNi$_2$O$_{7-\delta}$ &La$_2$NdNi$_2$O$_{7-\delta}$ \\ \hline
		$SJ_1$(meV) & 4.5 & 3.0 & 2.7 \\ \hline
		$SJ_1^{\prime}$(meV) & -3.4 & -3.5 & -3.5 \\ \hline
		$SJ_{\perp}$(meV) & 56.3 & 69.3 & 73.3 \\ \hline \hline
		$SJ_2$(meV) & 4.5 & 3.0 & 2.7 \\ \hline
		$SJ_2^{\prime}$(meV) & -3.6 & -3.6 & -3.6 \\ \hline
		$SJ_{\perp}$(meV) & 55.5 & 69.5 & 73.5 \\ \hline
	\end{tabular}
\end{table}

The variation in spin couplings may originate from the doping effect on the electronic properties of La$_3$Ni$_2$O$_{7-\delta}$, which introduces chemical pressure to change the orbital correlations\cite{pan2024,mzhang2024,gwang2025,huo2025,jgyang2024,yzhang2023c,zdong2025,cqchen2025}. The ionic radii of Pr and Nd are smaller than that of La, yielding a chemical pressure to compress the lattice by reducing the interlayer spacing and Ni-O-Ni bond angles(See Supplemental Materials). Such effect probably enhances the overlap between Ni-${d_{z^2}}$ and O-$p_z$ orbitals, but improves the critical pressure of superconductivity\cite{fyli2024a,fyli2025b,qzhong2025,jjfeng2024,zzqiu2025}. The antibonding state of the $d_{z^2}$ orbital shifts toward higher energy, which further strengthens the interlayer coupling. Meanwhile, the in-plane Ni-O bond length changes moderately,  exerting little influence on the in-plane orbital overlap. Therefore, the orbital-dependent effect from rare-earth doping is mainly concentrated on the $d_{z^2}$ orbital, giving a negligible impact on the electronic structure of the intralayer $d_{x^2-y^2}$ orbital. As a result, the orbital density and energy level distribution remain basically stable, and the intralayer coupling is almost unchanged overall. Since the ionic radius of Pr is slightly larger than that of Nd, the influence of the aforementioned effects is relatively weaker, and the corresponding variation is less pronounced, which is indeed consistent with our experimental data. If the interlayer $s\pm$-pairing indeed dominates under pressure, assuming $SJ_\perp \propto T_c$\cite{yyang2023,qqin2023,qqin2024,xzqu2024}, $SJ_\perp$= 73 meV would expected a promotion of $T_c$ to about 104 K in La$_2$NdNi$_2$O$_{7-\delta}$, surprisingly consistent with recent reports\cite{fyli2024a,fyli2025b,qzhong2025,jjfeng2024,zzqiu2025}. 

\section{Conclusions}
In conclusion, we have identified the spin excitations in the rare-earth doped bilayer nickelates La$_2$NdNi$_2$O$_{7-\delta}$ and La$_2$PrNi$_2$O$_{7-\delta}$. The flat feature of spin spectrum in La$_3$Ni$_2$O$_{7-\delta}$ keeps in Nd and Pr doped compounds, but it splits into two flat modes around 45 meV. By analyzing the spin excitations within the adopted nearest-neighbor stripe-type Heisenberg model, our results support an enhancement of the interlayer coupling $J_{\perp}$ in Nd and Pr doped samples, with estimated values in the range of about 69 to 73 meV. The stronger spin excitations and larger $J_{\perp}$ in La$_2$NdNi$_2$O$_{7-\delta}$ probably support higher $T_c$ under pressure than the parent compound La$_3$Ni$_2$O$_{7-\delta}$. Our results enlighten the mechanism investigations on the magnetically driven high-$T_c$ superconductivity in nickelates.

{\it This work is supported by the National Key Research and Development Program of China (Grants No.~2023YFA1406100, No.~2022YFA1403800 and No.~2025YFE0202100) and the National Natural Science Foundation of China (Grants No.~12574165, No.~12274444, and No.~12025408). DTA thank EPSRC-UK (Grant No. EP/W00562X/1) and the CAS for PIFI fellowship. Measurements on MERLIN were supported by the beam time allocation (Express proposals: 2590110 and 2590111) from the Science and Technology Facilities Council. Collected data from PANTHER (Proposal 4-04-536) are available at DOI:10.5291/ILL-DATA.4-04-536.}

\appendix

\begin{figure*}[htbp]
	\renewcommand\thefigure{S1}
	\includegraphics[width=0.95\textwidth]{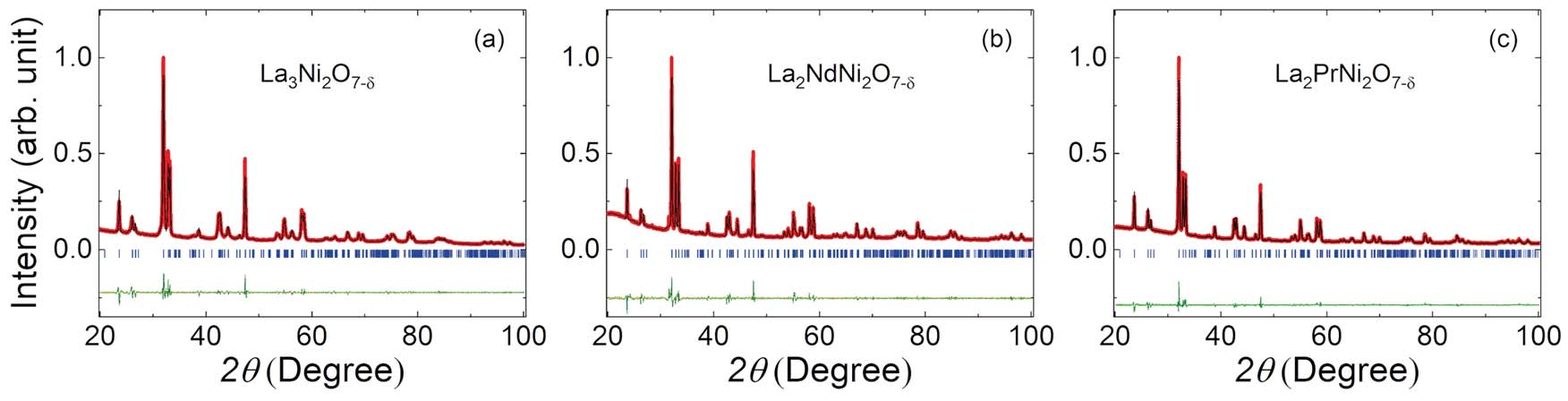}
	\caption{\label{S1}
		Powder X-ray diffraction and refinement results of (a) La$_3$Ni$_2$O$_{7-\delta}$ (b) La$_2$NdNi$_2$O$_{7-\delta}$ (c) La$_2$PrNi$_2$O$_{7-\delta}$ samples.
	}
\end{figure*}

\begin{center}
	{\bf Appendix A. SAMPLE CHARACTERIZATION}
\end{center}
The La$_3$Ni$_2$O$_{7-\delta}$, La$_2$NdNi$_2$O$_{7-\delta}$ and La$_2$PrNi$_2$O$_{7-\delta}$ polycrystalline samples were synthesized by the sol-gel method according to previous reports \cite{zhang1994,gwang2024,gwang2025}. The sample mass for each compound was about 5 grams. The phase purity and crystal structure of the our samples were determined by the powder X-ray diffraction (XRD) collected via an \emph{SmartLab} 9 kW diffractometer under high resolution mode at room temperature. The incident beam was set as Cu K$_\alpha$ radiation($\lambda$ = 1.5406 \AA), measurements were ranged from 20$^{\circ}$ to 100$^{\circ}$ for $2\theta$. Structural models are refined with the Rietveld method using the Fullprof software package and the refined results are shown in Fig. \ref{S1}. The refinement results of crystallographic data are listed in Table \ref{S1}. Although the weighted profile factor $R_{wp}$ is not so small for all three samples, all peaks can be indexed as the orthorhombic phase (space group $Amam$), while no impurity phases are found within the resolution of our XRD measurements. Clearly, the $c$-axis shrinks in Nd and Pr doped samples, both the out-of-plane (intralayer) Ni-O1-Ni and in-plane (interlayer) Ni-O3-Ni bonding angles decrease for smaller ionic radii of Nd and Pr, where the change of Ni-O bonding lengths is not very clear in doped samples, consistent with the previous reports on rare-earth doped bilayer nickelates\cite{gwang2025}.

\begin{table*}
	\renewcommand\thetable{S1}
	\caption{XRD refinement results of crystallographic data for three samples}
	\begin{ruledtabular}
		\begin{tabular}{cccc}
			Parameters& La$_3$Ni$_2$O$_{7-\delta}$ & La$_2$NdNi$_2$O$_{7-\delta}$ & La$_2$PrNi$_2$O$_{7-\delta}$\\
			\hline
			Space group & Amam & Amam & Amam\\
			R$_{wp}$ & 13.4 & 16.4 & 11.2   \\
			$\chi ^{2}$ & 2.33 & 1.75 & 1.37 \\
			a (\AA) & 5.400 & 5.367 & 5.376\\
			b (\AA) & 5.451 & 5.456 & 5.451\\
			c (\AA) & 20.500 & 20.357 & 20.387 \\
			Ni-O1-Ni angle (deg) & 168.35 & 165.49 & 162.10  \\
			Ni-O3-Ni angle (deg) & 171.65 & 166.73 & 164.64  \\
			Ni-O4-Ni angle (deg) & 167.76 & 168.60 & 166.04  \\
			Ni-O1 length (\AA) & 2.031 & 1.978 & 2.030  \\
			Ni-O2 length (\AA) & 2.169 & 2.198 & 2.096  \\
			Ni-O3 length (\AA) & 1.903 & 1.903 & 1.991  \\
			Ni-O4 length (\AA) & 1.950 & 1.947 & 1.871 \\
		\end{tabular}
	\end{ruledtabular}
\end{table*}

\begin{center}
	{\bf Appendix B. NEUTRON SCATTERING EXPERIMENTS}
\end{center}

Inelastic neutron scattering (INS) experiments on La$_2$NdNi$_2$O$_{7-\delta}$ were carried out using the MERLIN time-of-flight spectrometer at the ISIS Neutron and Muon Source, Rutherford Appleton Laboratory, UK\cite{bewley2006}. The incident energies were selected as $E_i=$ 15, 24, 50, 79 and 160 meV\cite{isisExp2025}. The INS spectra collected at MERLIN are already shown in Fig. 2 in the main text. Since the $E_i=$ 160 meV data could not be found any signals related to the spin excitations but only phonon and incoherent backgrounds, we do not present this data set in our paper. 

\begin{figure*}[htbp]
	\renewcommand\thefigure{S2}
	\includegraphics[width=0.95\textwidth]{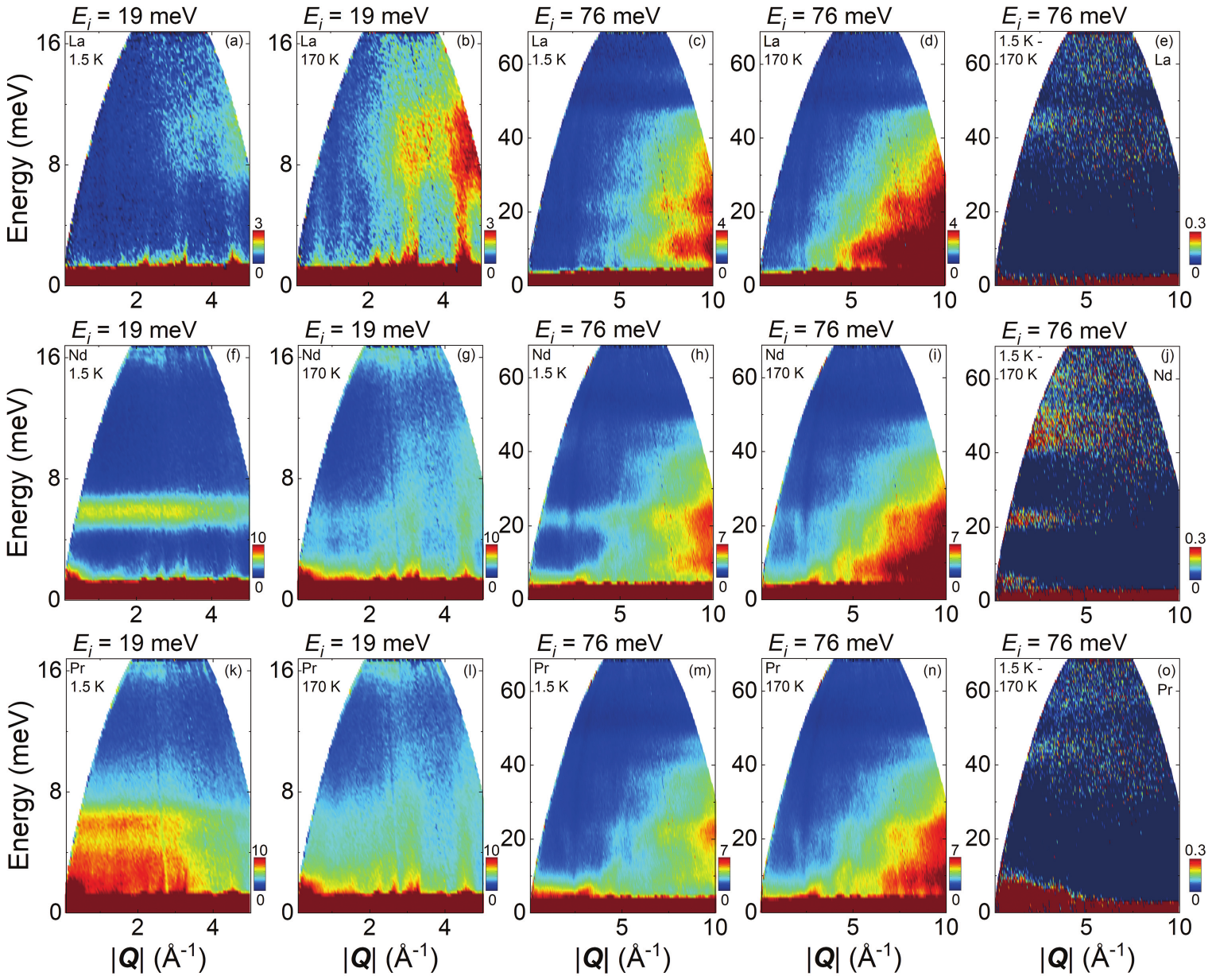}
	\caption{\label{S2}
		Raw data of INS spectra of La$_3$Ni$_2$O$_{7-\delta}$, La$_2$NdNi$_2$O$_{7-\delta}$ and La$_2$PrNi$_2$O$_{7-\delta}$ collected at PANTHER with $E_i=$ 19 and 76 meV. 
		For comparison, (e),(j) and (o)  present the subtracted spectra (1.5 K - 170 K) for all three samples without Bose corrections.
	}
\end{figure*}

\begin{figure*}[htbp]
	\renewcommand\thefigure{S3}
	\includegraphics[width=0.75\textwidth]{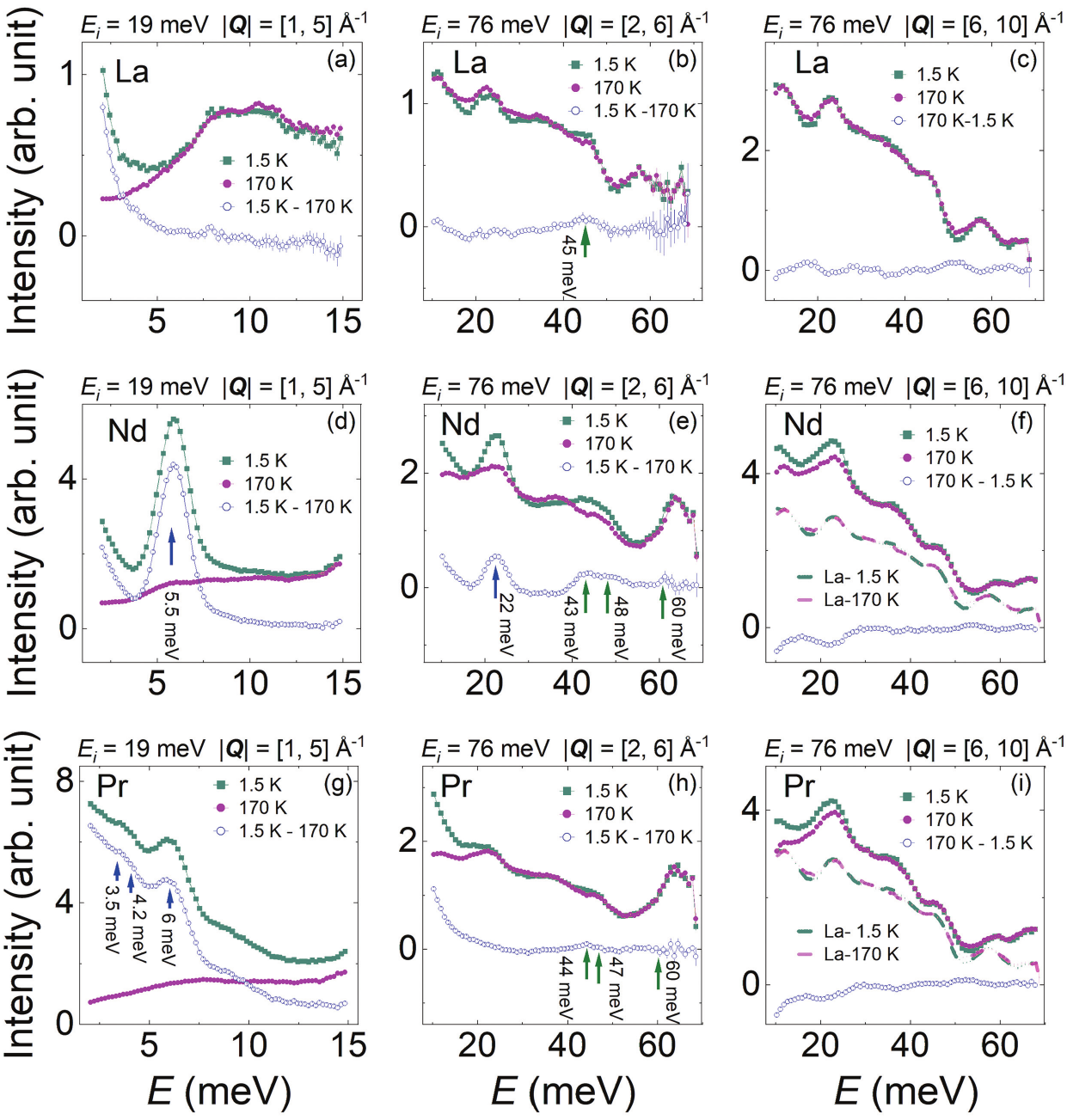}
	\caption{\label{S3}
		$E$-cuts at different regions $\mid\mathbf{Q}\mid$=[1, 5], [2, 6] and [6, 10] \AA$^{-1}$ for La$_3$Ni$_2$O$_{7-\delta}$, La$_2$NdNi$_2$O$_{7-\delta}$ and La$_2$PrNi$_2$O$_{7-\delta}$. All data are corrected by the Bose factor to easily compare between different temperatures. 
		The solid symbols represent the original data at $T=1.5$ and 170 K, and the open symbols are their differences. The blue arrows mark the CEF excitations, and the green arrows mark the spin excitations, respectively.
		The phonon excitations from the sample and aluminum can dominate at high $\mid\mathbf{Q}\mid$ range [6, 10] \AA$^{-1}$, where both Nd and Pr doped samples have similar phonon mode energies with pure La sample (dashed lines in (f) and (i)).
	}
\end{figure*}

\begin{figure*}[htbp]
	\renewcommand\thefigure{S4}
	\includegraphics[width=0.95\textwidth]{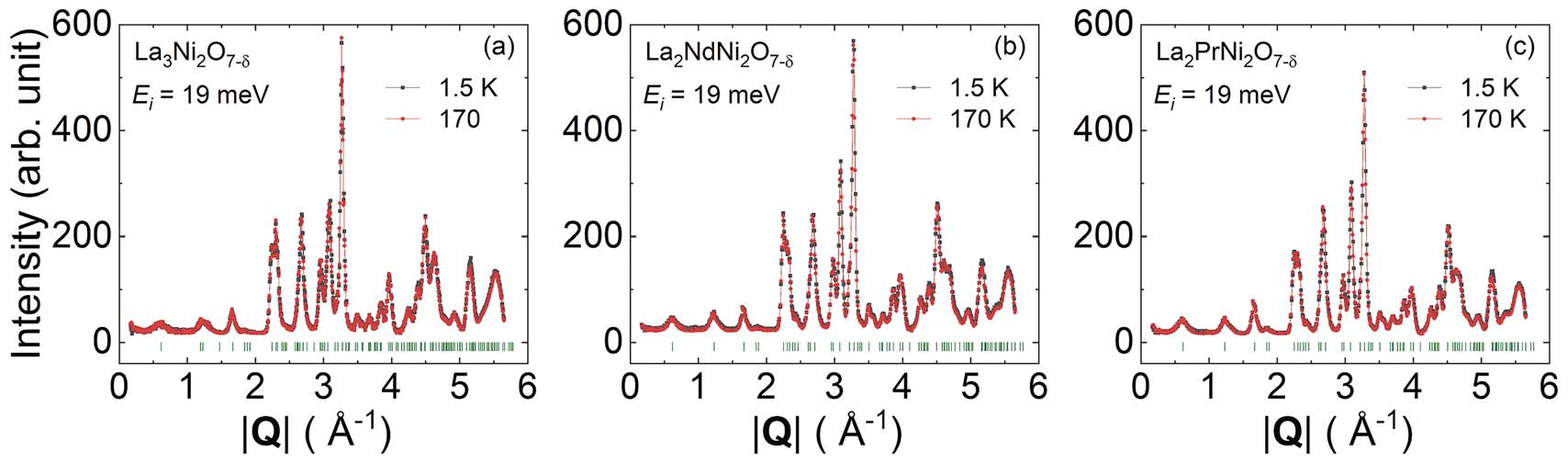}
	\caption{\label{S4}
		Quasi-elastic-cuts with $E$=[-0.5, 0.5] meV for (a) La$_3$Ni$_2$O$_{7-\delta}$, (b) La$_2$NdNi$_2$O$_{7-\delta}$ and (c) La$_2$PrNi$_2$O$_{7-\delta}$ collected at PANTHER. 
	}
\end{figure*}

Further INS measurements were performed at PANTHER time-of-flight spectrometer at the Institut Laue-Langevin, France\cite{Panther2022}. The incident energies were selected as $E_i=$ 19, and 76 meV for the measurements of La$_3$Ni$_2$O$_{7-\delta}$ and La$_2$NdNi$_2$O$_{7-\delta}$, and $E_i=$ 12.5, 19, and 76 meV for the measurements of La$_2$PrNi$_2$O$_{7-\delta}$\cite{illExp2025}. All three samples were measured at the base temperature $T=1.5$ K and high temperature $T=170$ K.  To track the temperature dependence of crystalline electric field (CEF) excitations, we also performed measurements at $T=50$ and 80 K for La$_2$PrNi$_2$O$_{7-\delta}$ sample. Fig. \ref{S2} presents the PANTHER data of INS spectra of La$_3$Ni$_2$O$_{7-\delta}$, La$_2$NdNi$_2$O$_{7-\delta}$ and La$_2$PrNi$_2$O$_{7-\delta}$ at $T=1.5$ and 170 K, along with their differences (Fig. \ref{S2} (e),(j) and (o)). It should be noted that we have plotted the intensity as arbitrary unit for convenience, and all subtraction between two temperatures are performed on the same sample, for their same conditions of absorption and self-shielding from the sample and similar background scattering from the empty can. However, since all PANTHER data are normalized by the scattering of an vanadium standard sample, the absolute intensity in the units of mbarn sr$^{-1}$ meV$^{-1}$ f.u.$^{-1}$ can be roughly obtained by dividing a factor of 10 from all plots in Fig. 3, Fig. 4, Fig. \ref{S2} and Fig. \ref{S3}, if it requires to compare with other systems. Clear CEF excitations can be observed at 5.5 and 22 meV for La$_2$NdNi$_2$O$_{7-\delta}$ (Fig. \ref{S2}(f) and (h)), while CEF signals in La$_2$PrNi$_2$O$_{7-\delta}$ are seen below 8 meV  (Fig. \ref{S2}(k)). Strong phonon signals dominate at the large momentum transfer range. After subtracting the data for high temperature (1.5 K - 170 K), La$_3$Ni$_2$O$_{7-\delta}$ shows weak flat-mode spin excitations around 45 meV (Fig. \ref{S2}(e)), consistent with the results of previous report\cite{txie2024}.  Splitting spin excitations around 45 meV can be observed both for La$_2$NdNi$_2$O$_{7-\delta}$ and La$_2$PrNi$_2$O$_{7-\delta}$ (Fig. \ref{S2}(j) and (o)).  Here at such high energy of spin excitations, the correction of Bose population factor $f_B=[1 - \exp(-\hbar\omega/k_B T)]$ is nearly negligible according to the fluctuation-dissipation theorem. Fig. \ref{S3} presents the 1D energy cuts at different $\mid\mathbf{Q}\mid$ ranges in Fig. \ref{S2}, where the signals of La$_2$NdNi$_2$O$_{7-\delta}$ and La$_2$PrNi$_2$O$_{7-\delta}$ are much stronger than those in La$_3$Ni$_2$O$_{7-\delta}$. From the difference between 1.5 K and 170 K after correcting by the Bose factor, CEF and spin excitations can be identified by peak features. Indeed, the spin excitation at 45 meV of La$_3$Ni$_2$O$_{7-\delta}$ splits into 43, 48 meV and 44, 47 meV in La$_2$NdNi$_2$O$_{7-\delta}$ and La$_2$PrNi$_2$O$_{7-\delta}$ as shown in the Fig. \ref{S3}(e) and (h), respectively. To compare the phonon intensities, we also perform the 1D cut with $\mid\mathbf{Q}\mid=$ [6, 10] \AA\ for $E_i=76$ meV (Fig. \ref{S3}(c),(f) and (i)), although the phonon spectra of Nd and Pr doped samples are slightly stronger than the pure La sample, the overall feature and mode energies are the same. It should be noted that the phonon background from the aluminum sample can is also included in such high $\mid\mathbf{Q}\mid$ range. The phonon background becomes similar at low $\mid\mathbf{Q}\mid$ range for all three samples\cite{txie2024,jyyou2025}, while electron-phonon coupling alone is insufficient to trigger superconductivity in La$_3$Ni$_2$O$_{7-\delta}$ under pressure\cite{jzhan2025}. Therefore, it is reasonable to search the spin excitations by subtracting the raw data of La$_3$Ni$_2$O$_{7-\delta}$ from those results of La$_2$NdNi$_2$O$_{7-\delta}$ and La$_2$PrNi$_2$O$_{7-\delta}$. Quasi-elastic-cuts with $E$=[-0.5, 0.5] meV for all three sample are presented in Fig. \ref{S4}, at $T=$ 1.5 K, no additional peaks or superlattice reflections are observed compared to the high-temperature (170 K) data. These quasi-elastic neutron scattering data do not provide any evidences for the formation of a spin density wave. However, since PANTHER is a spectrometer for inelastic neutron scattering, diffraction measurements will be more suitable for the exploration on the possible static magnetic orders.

\begin{center}
	{\bf Appendix C. LINEAR SPIN-WAVE THEORY CALCULATIONS}
\end{center}

To determine the specific magnitudes of the magnetic exchange couplings, we used the SpinW software package and calculated the dispersion relations of the double spin stripe (DSS), single spin-charge stripe (SCS), A-type antiferromagnetic order (AFM-A), and G-type antiferromagnetic order (AFM-G) based on the Linear Spin-Wave Theory(LSWT).

The basic procedure for calculating the spin dynamical dispersion relations is shown as follows: First, we consider the Hamiltonian by using the Heisenberg model
\begin{equation}
	H=\sum_{<i,j>} J_{ij}{S_{i}}\cdot {S_{j}},
\end{equation}
where $S_i$ denotes the spin operator on each site $i$, ${\left\langle {i,j}\right\rangle}$ denotes neighbor sites. $J_{ij}$ is the exchange interactions between $i$th and $j$th
spins. Using the commutation relations of angular momentum operators and introducing $S^{\pm}$, we have
\begin{equation}
	H=\sum_{<i,j>} J_{ij}[{S_{i}^{z}}\cdot {S_{j}^{z}}+\frac{1}{2}({S_{i}^{+}}\cdot {S_{j}^{-}}+{S_{i}^{-}}\cdot {S_{j}^{+}})],
\end{equation}
where$S_{i}^{\pm } =S_{i}^{x}\pm  S_{i}^{y}$. To further diagonalize the Hamiltonian, we perform the Holstein-Primakoff transformation, introduce spin deviation quantum number $n_{i}=S-m_{i}$ as well as its corresponding creation and annihilation operators $a_{i}$

\begin{equation}
	\left\{\begin{matrix}
		a_{i}\left | n_{i}  \right \rangle =\sqrt{n_{i}} \left | n_{i}-1  \right \rangle  \\
		a_{i}^{\dagger} \left | n_{i}  \right \rangle =\sqrt{n_{i}+1} \left | n_{i}+1  \right \rangle  \\
		a_{i}^{\dagger}a_{i} \left | n_{i}  \right \rangle =\sqrt{n_{i}} \left | n_{i}  \right \rangle\\
	\end{matrix}\right.
\end{equation}
and perform a Fourier transform on $a_{i}$
\begin{equation}
	a_{i}=N^{-\frac{1}{2}}\sum_{k}e^{ik\cdot R_{l}}a_{k},a_{i}^{\dagger }=N^{-\frac{1}{2}}\sum_{k}e^{ik\cdot R_{l}}a_{k}^{\dagger }.
\end{equation}
From this, we can obtain the Hamiltonian under the low-temperature approximation
\begin{equation}
	H=E_{Cl}+\sum_{k,\left\langle {i,j} \right\rangle} C_{ii}a_{k,i}^{\dagger}a_{k,i}+C_{ij}(a_{k,i}a_{k,j}+a_{k,i}^{\dagger}a_{k,j}^{\dagger}),
\end{equation}
where $E_{Cl}$ is the ground state energy. It should be noted that the indices $i$ and $j$ here do not refer to the $i$th or $j$th lattice sites in real space, but to the $a_{k}$ operators corresponding to different lattice sites within a unit cell. If there are a total of $i$ lattice sites in a unit cell, there will be $i$ corresponding $a_{k}$ operators. $C_{ii},C_{ij}$ are coefficients related to the specific lattice point in the unit cell.
We diagonalize the Hamiltonian using the extended Bogoliubov transformation
\begin{equation}
	b_{k,i}=\sum_{j} m_{ij}a_{k,j}+m'_{ij}a_{k,j}^{\dagger}.
\end{equation}
By substituting the transformed Hamiltonian into the quantum equation of motion
\begin{equation}
	i\hbar \dot{b}_{k,i}=[b_{k,i}, H], 
\end{equation}
the spin dynamical dispersion relations can be obtained accordingly. Considering the simplicity of the model, only the nearest-neighbor interactions were taken into account in our calculations. Therefore, the Hamiltonian expression of the DSS state could be presented as:
\begin{equation}
	H=J_{1}\sum_{<i,j>} {S_{i}}\cdot {S_{j}}+J_{1}'\sum_{<i,j>'} {S_{i}}\cdot {S_{j}}+J_{\bot }\sum_{i} {S^{t}_{i}}\cdot {S^{b}_{i}},
\end{equation}
where $J_1$ and $J_1^{\prime}$ are the antiferromagnetic (AF) and ferromagnetic (FM) intralayer nearest-neighbor interactions, $J_\perp$ is the interlayer nearest-neighbor interaction, $<i,j>$ and $<i,j>^{\prime}$ denote the intralayer nearest-neighbor lattice sites, and $S^{t}$, $S^{b}$ denotes the spins at the nearest-neighbor lattice site for the top and bottom layers.

\begin{figure*}[htbp]
	\renewcommand\thefigure{S5}
	\includegraphics[width=0.95\textwidth]{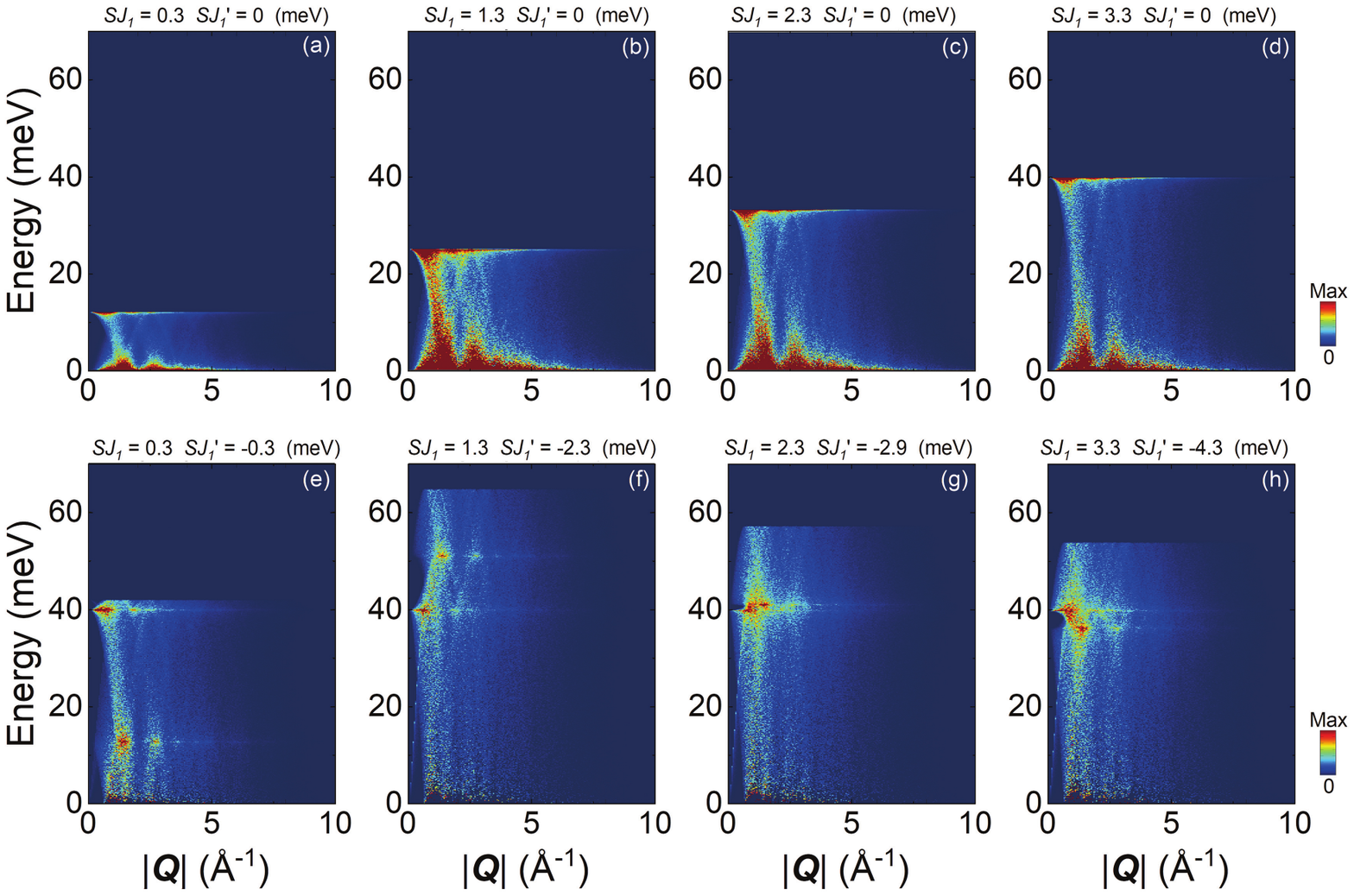}
	\caption{\label{S5}
		SpinW calculations based on the double spin stripe (DSS) by fixing $SJ_{\perp}$ = 60 meV. 
		(a)-(d) SpinW calculations by only considering the nearest-neighbor intralayer AF exchange coupling $J_1$.
		(e)-(h) SpinW calculations by considering both the nearest-neighbor AF-type intralayer exchange coupling $J_1$ and the nearest-neighbor FM-type exchange coupling $J_1^{\prime}$.
	}
\end{figure*}

\begin{figure*}[htbp]
	\renewcommand\thefigure{S6}
	\includegraphics[width=0.75\textwidth]{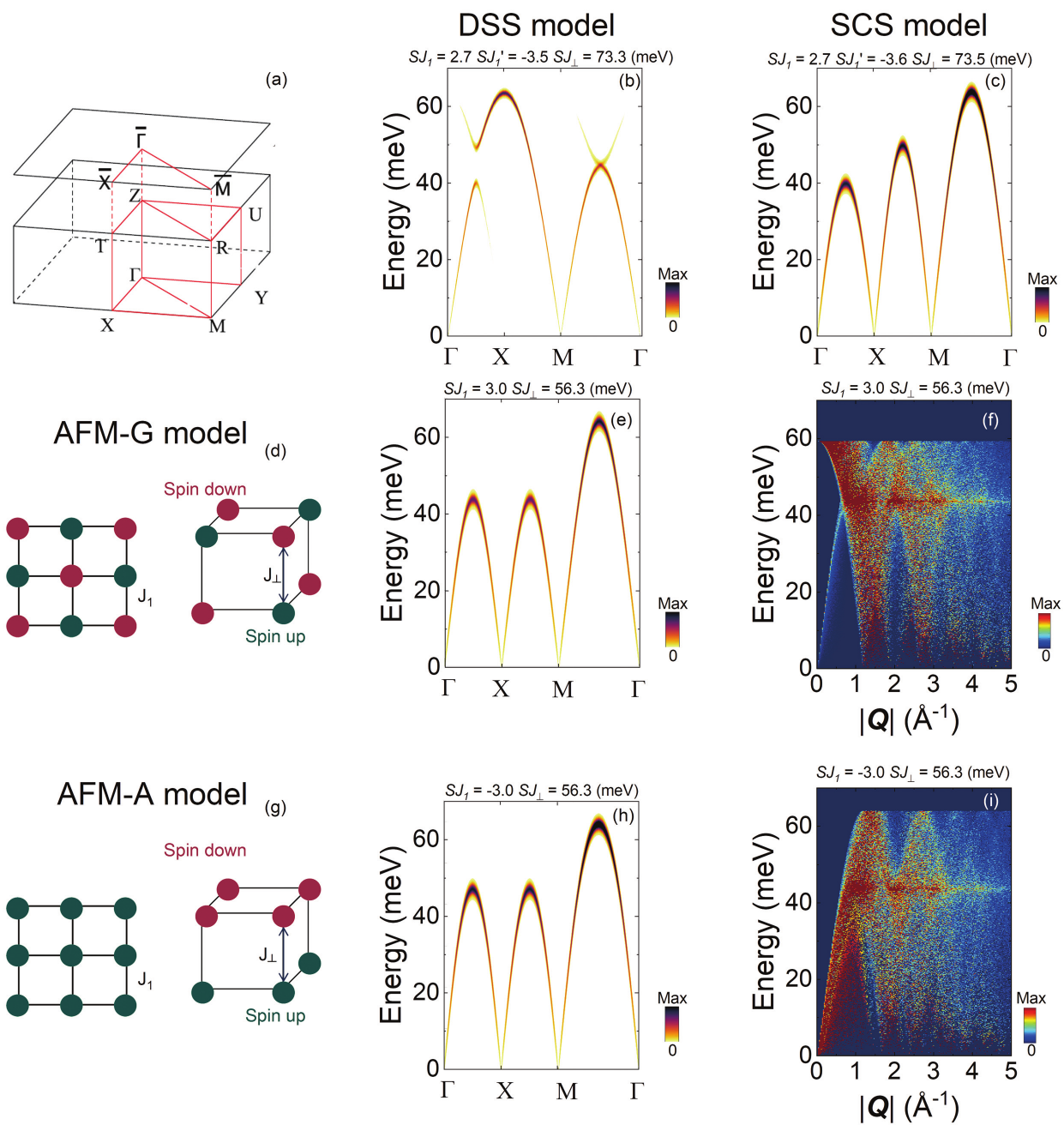}
	\caption{\label{S6}
		Spin wave dispersions and magnetic configurations in our SpinW models.
		(a) Brillouin zone of La$_3$Ni$_2$O$_{7-\delta}$.
		(b) and (c) Spin wave dispersions obtained from fitting parameters of exchange couplings for DSS and SCS models listed in Table. 1 and Fig. 1.
		(d)-(f) Spin configuration, exchange couplings, dispersion and spectrum for G-type AF model (AFM-G) with $SJ_1$= 3.0 meV and $SJ_{\perp}$ = 56.3 meV.
		(d)-(f) Spin configuration, exchange couplings, dispersion and spectrum for A-type AF model (AFM-A) with $SJ_1$= -3.0 meV and $SJ_{\perp}$ = 56.3 meV.
	}
\end{figure*}

By fixing $SJ_\perp = 60 $ meV and setting $SJ_1^{\prime} = 0$, it can be obtained that when $SJ_1$ varies within the range of $0.3 \sim 3.3$ meV, a distinct flat mode forms in the excitation spectrum of powder sample (Fig.\ref{S5} (a)-(d)). Meanwhile, the larger value of $SJ_1$, the higher energy of this mode. This confirms that the prominent flat-band excitation spectrum observed in experiments arises from the interlayer interaction, which is much stronger than the intralayer interaction\cite{xie2024}. When FM-type $SJ_1^{\prime}$ is further introduced while keeping $SJ_\perp= 60$ meV unchanged, adjusting the magnitude of the intralayer nearest-neighbor interaction can account for the splitting excitation spectrum around 45 meV in the rare-earth doped compounds (Fig.\ref{S5} (e)-(h)).

Similarly, the Hamiltonian expression of the SCS model is shown as follows:
\begin{equation}
	H=J_{2}\sum_{<i,j>} {S_{i}}\cdot {S_{j}}+J_{2}'\sum_{<i,j>'} {S_{i}}\cdot {S_{j}}+J_{\bot }\sum_{i} {S^{t}_{i}}\cdot {S^{b}_{i}}.
\end{equation}
Here, $J_2$ and $J_2^{\prime}$ denote the AF and FM intralayer nearest-neighbor interactions along the diagonal directions in the orthorhombic lattice, respectively. We present the calculated dispersion relations for DSS and SCS models in Fig.\ref{S6} (b) and (c). The splitting flat excitations originate either from the band gap in between $\Gamma$ and X points for DSS model, or from the different bandtops between the paths of $\Gamma$ to X and X to M points. Further INS measurements on single-crystalline samples are highly desired to resolve this issue.

In contrast, the Hamiltonians of the AFM-A and AFM-G models can be uniformly expressed as follows:
\begin{equation}
	H=J_{1}\sum_{<i,j>} {S_{i}}\cdot {S_{j}}+J_{\bot }\sum_{i} {S^{t}_{i}}\cdot {S^{b}_{i}},
\end{equation}
where for AFM-A, $J_1$ is an antiferromagnetic interaction, and for AFM-G, $J_1$ is a ferromagnetic interaction.
Setting $SJ_1= 3$ meV and $SJ_\perp = 60$ meV, it can be observed that the excitation spectrum exhibits wave-like characteristics rather than flat-band characteristic even for powder sample (Fig.\ref{S6} (d)-(i)), which is inconsistent with the experimental results. Furthermore, it cannot explain the splitting observed in La$_2$NdNi$_2$O$_{7-\delta}$ and La$_2$PrNi$_2$O$_{7-\delta}$, since the bandtops between $\Gamma$ to X and X to M points are always the same. Therefore, our experimental results can rule out the possibility of AFM-A and AFM-G models in the bilayer nickelate system.

It should be particularly noted that the exchange parameter \(J\) referred to in this work denotes the effective exchange parameter determined via precise diagonalization combined with Heisenberg model mapping, which inherently incorporates the contribution of energy-level splitting induced by the CEF effect from Ni$^{n+}$ ions. The total Hamiltonian of the multi-orbital transition metal oxide system we constructed is the core basis for calculating the effective exchange interactions, which explicitly includes the crystal electric field splitting term ($H_{\mathrm{CEF}}$), orbital hopping term and Hubbard Coulomb repulsion term ($H_U$):
\begin{equation}
	H=\sum_{\sigma }(\epsilon _dn_{d\sigma }+\epsilon _pn_{p\sigma }-(t_{pd}c_{d\sigma }^\dagger c_{p\sigma }+h.c.))+U_d\sum_{i}n_{i\uparrow }n_{i\downarrow }.
\end{equation}
Here, the crystal field term is given by $H_{\mathrm{CEF}}=\sum_{\sigma }(\epsilon _dn_{d\sigma }+\epsilon _pn_{p\sigma })$, where $\epsilon _\alpha $ denotes the crystal field-induced energy splitting of different d-orbitals [e.g., the energy difference between the $e_g$ orbitals ($d_{z^2}$,$d_{x^2-y^2}$ ) and $t_{2g}$ orbitals ($d_{xy}$,$d_{xz}$,$d_{yz}$) of Ni$^{n+}$]. For the nickelate oxides investigated herein, the crystal field splits the 3d orbitals of Ni into the higher-energy $e_g$ orbitals and lower-energy $t_{2g}$ orbitals; additionally, the $e_g$ orbitals undergo further splitting under tetragonal distortion (coordination of apical oxygen and in-plane oxygen), a characteristic that is fully embodied in the parameter of the Hamiltonian. The orbital hopping term is expressed as $H_{\mathrm{hop}}=\sum_{\sigma }(t_{pd}c_{d\sigma }^\dagger c_{p\sigma }+h.c.)$, which includes the hopping between Ni d-orbitals and O p-orbitals ($t_{pd}$). From this, the Hamiltonian expressions for the triplet and singlet states can be derived respectively as:
\begin{equation}
	^{Tri}H_{eff}=-\frac{2\left | t_{pd} \right | ^2}{U+\Delta _{pd}},
\end{equation}
\begin{equation}
	^{Sin}H_{eff}=\\
	-\frac{2\left | t_{pd} \right | ^2}{U+\Delta _{pd}}\begin{bmatrix}
		1& 0\\
		0&1
	\end{bmatrix} \\
			-\left | t_{pd} \right |^4 
			\left \{ (\frac{1}{U}+1 )\frac{1}{(U+\Delta _{pd})^2} \right \} \begin{bmatrix}
		1& -1\\
		-1&1
	\end{bmatrix} ,
\end{equation}
Thus, we can obtain the super-exchange interaction integral J between two identical d-orbitals as:
\begin{equation}
	J_{eff}=\frac{4t_{pd}^4}{(U+\Delta _{pd})^2} (\frac{1}{U}+\frac{1}{U+\Delta _{pd}}  )
\end{equation}

In summary, the effective exchange parameter calculated in our original work is derived from the eigenenergy of the full Hamiltonian including the CEF splitting term, and the CEF-induced energy level splitting effect is fully incorporated into the calculation through the exact diagonalization of the multi-orbital cluster model and Heisenberg model mapping. After considering the joint effects of CEF, orbital hopping and Coulomb repulsion, the obtained parameter \(J\) are physically meaningful for describing the magnetic correlation in nickelates.

\begin{center}
	{\bf Appendix D. CEF MODEL OF RARE-EARTH ELEMENTS}
\end{center}

In the 4$f$-electron rare earth systems, the magnetic ions interact with the surrounding CEF of the ligand atoms based on the single-ion model \cite{Fulde1985,Mackintosh1992}. The magnetism in such systems originates from the total angular momentum of the 4$f$ electrons. Above the magnetic ordering temperature, the single-ion model can effectively reproduce both the CEF excitations observed by the measurements from INS and thermodynamic properties, such as the temperature dependence of the magnetic heat capacity and entropy. The allowed single-ion terms are strongly restricted by the space group symmetry of the crystal structure. Specifically, $\mathrm{La}_{2} \mathrm{PrNi}_{2} \mathrm{O}_{7-\delta}$ and $\mathrm{La}_{2} \mathrm{NdNi}_{2} \mathrm{O}_{7-\delta}$ crystallize in the space group Amam (No. 63) with the doped rare-earth atoms occupying two distinct Wyckoff positions: $4 c$ (site symmetry $m 2 m$, corresponding to $C_{2 v}$ point symmetry) and $8 g$ (site symmetry ..$m$, corresponding to $C_{s}$ point symmetry). Considering the point symmetry $C_{2 v}$ and $C_{s}$ of the magnetic ions, the CEF Hamiltonians with the quantization axis along the c axis are given by:

\begin{equation}
	\begin{aligned}
		\widehat{H}_{C E F}^{C_{2 v}} = & B_{2}^{0} \widehat{O}_{2}^{0}+B_{2}^{2} \hat{O}_{2}^{2} \\
		& +B_{4}^{0} \widehat{O}_{4}^{0}+B_{4}^{2} \hat{O}_{4}^{2}+B_{4}^{4} \hat{O}_{4}^{4} \\
		& +B_{6}^{0} \hat{O}_{6}^{0}+B_{6}^{2} \hat{O}_{6}^{2}+B_{6}^{4} \hat{O}_{6}^{4}+B_{6}^{6} \hat{O}_{6}^{6} ,
	\end{aligned}
\end{equation}

and\\

\begin{equation}
	\begin{aligned}
		\widehat{H}_{C E F}^{C_{S}} = & B_{2}^{0} \widehat{O}_{2}^{0}+B_{2}^{1} \widehat{O}_{2}^{1}+B_{2}^{2} \widehat{O}_{2}^{2}\\
		& +B_{4}^{0} \hat{O}_{4}^{0}+B_{4}^{1} \hat{O}_{4}^{1}+B_{4}^{2} \hat{O}_{4}^{2}+B_{4}^{3} \hat{O}_{4}^{3}+B_{4}^{4} \hat{O}_{4}^{4} \\
		& +B_{6}^{0} \hat{O}_{6}^{0}+B_{6}^{1} \hat{O}_{6}^{1}+B_{6}^{2} \hat{O}_{6}^{2}+B_{6}^{3} \hat{O}_{6}^{3}+B_{6}^{4} \hat{O}_{6}^{4}+B_{6}^{5} \hat{O}_{6}^{5}+B_{6}^{6} \hat{O}_{6}^{6}, 
	\end{aligned}
\end{equation}
respectively. Here, $B_{m}^{n}$ and $\widehat{O}_{m}^{n}$ are Stevens parameters and Stevens operators, respectively \cite{Rhyne1967,Hutchings1964}. The degeneracy of the total angular momentum $J$ is split into several CEF levels once we consider these single-ion Hamiltonians, and the matrix elements between these states can be easily evaluated. At a given wave vector $\boldsymbol{Q}$, the neutron cross section associated with the CEF Hamiltonian is \cite{Boothroyd2020}:

\begin{equation}
	\frac{d^{2} \sigma}{d \Omega d \omega}=A \sum_{m, n} p_{n}|\langle\Gamma_{m}| \hat{J}_{\perp}| \Gamma_{n}\rangle|^{2} \delta(\hbar \omega+E_{n}-E_{m}),
\end{equation}
where $A$ is a normalization factor, $p_{n}$ is the Boltzmann weight, $\hat{J}_{\perp}$ is the component of $\hat{J}$ perpendicular to $\boldsymbol{Q}$, and $|\langle\Gamma_{m}| \hat{J}_{\perp}| \Gamma_{n}\rangle|^{2}$ is calculated from the inner product of the matrix element of the magnetic moment with the CEF eigenstates $\left|\Gamma_{n}\right\rangle$.

To better understand the CEF level scheme, the magnetic entropy ($S_{m}$) is derived, as shown in FIG. 1($l$), by integrating $C_{m} / T$ with respect to temperature $(T)$. The magnetic heat capacity ($C_{m}$) of $\mathrm{La}_{2} \mathrm{RENi}_{2} \mathrm{O}_{7-\delta}(\mathrm{RE}=\mathrm{Pr}, \mathrm{Nd})$ was obtained by subtracting the phonon background and the magnetic contribution from Ni magnon in the isostructural $\mathrm{La}_{3} \mathrm{Ni}_{2} \mathrm{O}_{7-\delta}$. It is worth noting that the magnetic entropy of these two samples exceeds the plateau expected from the CEF contribution of $\operatorname{Pr}^{3+}\left(4 f^{2}, J=4\right)$ and $\mathrm{Nd}^{3+}\left(4 f^{3}, J=9 / 2\right)$ atoms. This excess entropy suggests the presence of enhanced spin fluctuations of Ni in $\mathrm{La}_{2} \mathrm{RENi}_{2} \mathrm{O}_{7}$ ${ }_{\delta}$ compared to $\mathrm{La}_{3} \mathrm{Ni}_{2} \mathrm{O}_{7-\delta}$.

The CEF excitations were fitted simultaneously using a least-squares minimization approach based solely on the neutron scattering data, employing the PYCRYSTALFIELD CEF calculation package \cite{Scheie2021}. Here we specify the CEF fitting data in different materials in the following subsections.

\subsection{$\mathrm{La}_{2} \mathrm{NdNi}_{2} \mathrm{O}_{7-\delta}$}

\begin{figure}[htbp]
	\renewcommand\thefigure{S7}
	\includegraphics[width=0.45\textwidth]{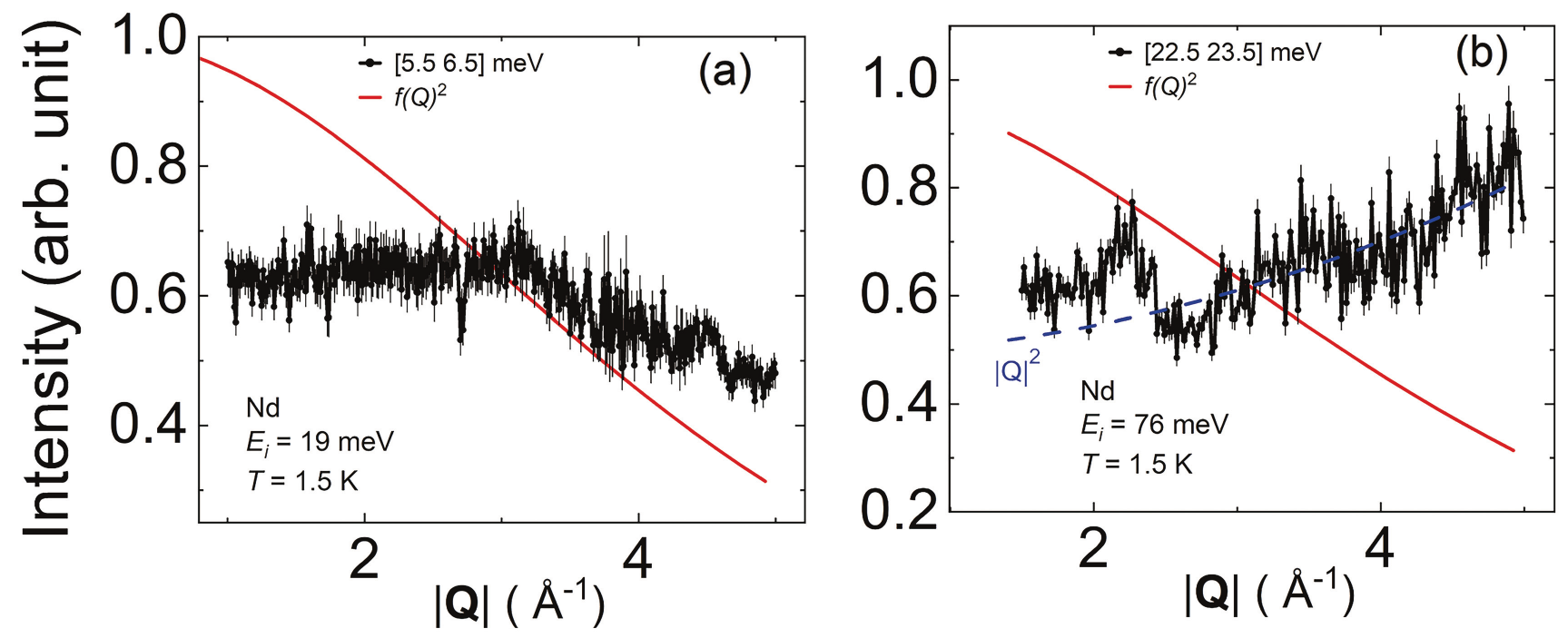}
	\caption{\label{S7}
		$Q$-cuts of the $\mathrm{La}_{2} \mathrm{NdNi}_{2} \mathrm{O}_{7-\delta}$ data with (a) $E_{i}=19$ meV and (b) $E_{i}=76$ meV  integrated over listed energy ranges. The red solid lines are the square of the magnetic form factor for $\mathrm{Nd}^{3+}$, and the blue line is the square of momentum transfer $|\boldsymbol{Q}|^{\mathbf{2}}$ after normalizing by the intensity.
	}
\end{figure}

The energy cuts for $\mathrm{La}_{2} \mathrm{NdNi}_{2} \mathrm{O}_{7-\delta}$ from $E_i=19$ meV and 76 meV data set reveal two different CEF modes: 6 meV and 23 meV. It should be noted that they are slightly different from those peaks at 5.5 meV and 22 meV obtained from the 1.5 K - 170 K data, due to the peak shape changing with temperature (Fig.  S2 and S3). Integrating over a finite energy range in the relevant spectra allows us to compare the $|\boldsymbol{Q}|$ dependence of these excitations with the expectations for magnetic and phonon modes. For the 6 meV mode, the $Q$-cut with 5.5 meV $\leq E \leq$ 6.5 meV at high $|\boldsymbol{Q}|$ shown in FIG.\ref{S7}(a) has $\boldsymbol{I} \propto$ $f(|\boldsymbol{Q}|)^{2}$, where $f(|\boldsymbol{Q}|)$ is the magnetic form factor for $\mathrm{Nd}^{3+}$, and therefore these excitations are identified as CEF levels of $\mathrm{Nd}^{3+}$. However, the $Q$-cut with 22.5 meV $\leq E \leq$ 23.5 meV reveals the modes at $E=23$ meV mixed with phonon signals at high $\boldsymbol{Q}$, as the $|\boldsymbol{Q}|$ dependence of intensity varies as $\boldsymbol{I} \propto |\boldsymbol{Q}|^{\mathbf{2}}$ instead of $f(|\boldsymbol{Q}|)^{2}$ (FIG. \ref{S7}(b)). Starting from these two CEF modes, we tried to fit the CEF paprameters.

Using Stevens-operator equivalent method, we separately carried out CEF fitting for two different Wyckoff positions of $\mathrm{Nd}^{3+}$ \cite{critter2021,ryamamoto2023,lyang2025,hnair2017}. The fitting results and eigenvalues and eigenvectors are shown in FIG.\ref{S9} (a)-(d) (red and blue lines), Table \ref{tableS2} and \ref{tableS3}, respectively. The CEF parameters of $\mathrm{La}_{2} \mathrm{NdNi}_{2} \mathrm{O}_{7-\delta}$ with $\mathrm{Nd}^{3+}$ at $4 c$ Wyckoff positions are $B_{2}^{0}=4.477 \times 10^{-2} \mathrm {meV}$, $B_{2}^{2}=3.132 \times 10^{-2} \mathrm {meV}, ~ B_{4}^{0}=4.257 \times 10^{-3} \mathrm {meV}, B_{4}^{2}=1.805 \times 10^{-3} \mathrm {meV}, B_{4}^{4}= 2.983 \times 10^{-2} \mathrm {meV}$, $B_{6}^{0}=-7.422 \times 10^{-6} \mathrm {meV}$, $B_{6}^{2}=2.927 \times 10^{-5} \mathrm {meV}$, $B_{6}^{4}=-5.987 \times 10^{-4} \mathrm {meV}, B_{6}^{6}=-2.345 \times 10^{-4} \mathrm {meV}$. The CEF parameters of $\mathrm{La}_{2} \mathrm{NdNi}_{2} \mathrm{O}_{7-\delta}$ with $\mathrm{Nd}^{3+}$ at 4c Wyckoff positions are $B_{2}^{0}=-2.933 \times 10^{-1} \mathrm {meV}$, $B_{2}^{1}=6.806 \times 10^{-2} \mathrm {meV}$, $B_{2}^{2}= 1.681 \times 10^{-3} \mathrm {meV}, ~ B_{4}^{0}=8.243 \times 10^{-4} \mathrm {meV}, ~ B_{4}^{1}=-1.841 \times 10^{-3} \mathrm {meV}, ~ B_{4}^{2}=5.823 \times 10^{-5} \mathrm {meV}, B_{4}^{3}=1.157 \times 10^{-3} \mathrm {meV}, B_{4}^{4}=2.1402 \times 10^{-2} \mathrm {meV}, B_{6}^{0}=-2.116 \times 10^{-5} \mathrm {meV}$, $B_{6}^{1}=-2.234 \times 10^{-4} \mathrm {meV}, ~ B_{6}^{2}=-3.946 \times 10^{-5} \mathrm {meV}, ~ B_{6}^{3}=8.487 \times 10^{-4} \mathrm {meV}, B_{6}^{4}= -4.066 \times 10^{-4} \mathrm {meV}, B_{6}^{5}=2.884 \times 10^{-3} \mathrm {meV}, B_{6}^{6}=-1.4647 \times 10^{-4} \mathrm {meV}$.

\begin{table*}
	\renewcommand\thetable{S2}
	\caption{Energy levels and associated wave functions determined from the analysis of the INS data of $\mathrm{La}_{2} \mathrm{NdNi}_{2} \mathrm{O}_{7-\delta}$ with Nd at $4 c$ Wyckoff positions using the $C_{2 v}$ CEF model described in the text.}
	\begin{ruledtabular}
		\begin{tabular}{c|cccccccccc}
			E (meV) &$| -\frac{9}{2}\rangle$ & $| -\frac{7}{2}\rangle$ & $| -\frac{5}{2}\rangle$ & $| -\frac{3}{2}\rangle$ & $| -\frac{1}{2}\rangle$ & $| \frac{1}{2}\rangle$ & $| \frac{3}{2}\rangle$ & $| \frac{5}{2}\rangle$ & $| \frac{7}{2}\rangle$ & $| \frac{9}{2}\rangle$ \tabularnewline
			\hline 
			0.000 & 0.0 & -0.1918 & 0.0 & 0.5772 & 0.0 & 0.0586 & 0.0 & -0.7897 & 0.0 & 0.0545 \tabularnewline
			0.000 & 0.0545 & 0.0 & -0.7897 & 0.0 & 0.0586 & 0.0 & 0.5772 & 0.0 & -0.1918 & 0.0 \tabularnewline
			5.996 & 0.0 & 0.9095 & 0.0 & 0.1337 & 0.0 & -0.3641 & 0.0 & -0.1491 & 0.0 & 0.0148 \tabularnewline
			5.996 & -0.0148 & 0.0 & 0.1491 & 0.0 & 0.3641 & 0.0 & -0.1337 & 0.0 & -0.9095 & 0.0 \tabularnewline
			22.965 & 0.609 & 0.0 & 0.2359 & 0.0 & 0.6568 & 0.0 & 0.2817 & 0.0 & 0.2503 & 0.0 \tabularnewline
			22.965 & 0.0 & -0.2503 & 0.0 & -0.2817 & 0.0 & -0.6568 & 0.0 & -0.2359 & 0.0 & -0.609 \tabularnewline
			23.035 & 0.0804 & 0.0 & 0.5166 & 0.0 & -0.4728 & 0.0 & 0.6788 & 0.0 & -0.2057 & 0.0 \tabularnewline
			23.035 & 0.0 & 0.2057 & 0.0 & -0.6788 & 0.0 & 0.4728 & 0.0 & -0.5166 & 0.0 & -0.0804 \tabularnewline
			23.047 & 0.787 & 0.0 & -0.1778 & 0.0 & -0.4572 & 0.0 & -0.3298 & 0.0 & -0.1765 & 0.0 \tabularnewline
			23.047 & 0.0 & 0.1765 & 0.0 & 0.3298 & 0.0 & 0.4572 & 0.0 & 0.1778 & 0.0 & -0.787 \tabularnewline
	\end{tabular}\end{ruledtabular}
	\label{tableS2}
\end{table*}

\begin{table*}
	\renewcommand\thetable{S3}
	\caption{Energy levels and associated wave functions determined from the analysis of the INS data of $\mathrm{La}_{2} \mathrm{NdNi}_{2} \mathrm{O}_{7-\delta}$ with Nd at $8 g$ Wyckoff positions using the $C_{s}$ CEF model described in the text.}
	\begin{ruledtabular}
		\begin{tabular}{c|cccccccccc}
			E (meV) &$| -\frac{9}{2}\rangle$ & $| -\frac{7}{2}\rangle$ & $| -\frac{5}{2}\rangle$ & $| -\frac{3}{2}\rangle$ & $| -\frac{1}{2}\rangle$ & $| \frac{1}{2}\rangle$ & $| \frac{3}{2}\rangle$ & $| \frac{5}{2}\rangle$ & $| \frac{7}{2}\rangle$ & $| \frac{9}{2}\rangle$ \tabularnewline
			\hline 
			0.000 & 0.0 & -0.0364 & 0.1933 & 0.031 & -0.2221 & -0.0134 & -0.2574 & -0.0771 & 0.2214 & 0.8886 \tabularnewline
			0.000 & 0.8886 & -0.2214 & -0.0771 & 0.2574 & -0.0134 & 0.2221 & 0.031 & -0.1933 & -0.0364 & 0.0 \tabularnewline
			5.985 & 0.0494 & -0.1304 & -0.6554 & -0.3102 & 0.0324 & 0.1456 & 0.3464 & 0.4576 & -0.065 & 0.3146 \tabularnewline
			5.985 & 0.3146 & 0.065 & 0.4576 & -0.3464 & 0.1456 & -0.0324 & -0.3102 & 0.6554 & -0.1304 & -0.0494 \tabularnewline
			6.037 & 0.2589 & 0.8578 & -0.0368 & -0.0351 & -0.1417 & -0.1468 & 0.2792 & 0.0101 & 0.273 & 0.0205 \tabularnewline
			6.037 & -0.0205 & 0.273 & -0.0101 & 0.2792 & 0.1468 & -0.1417 & 0.0351 & -0.0368 & -0.8578 & 0.2589 \tabularnewline
			22.953 & 0.1751 & -0.293 & 0.0203 & -0.2081 & -0.1623 & -0.8707 & 0.2133 & -0.0925 & -0.0367 & 0.0 \tabularnewline
			22.953 & 0.0 & 0.0367 & -0.0925 & -0.2133 & -0.8707 & 0.1623 & -0.2081 & -0.0203 & -0.293 & -0.1751 \tabularnewline
			23.043 & -0.1043 & -0.1687 & 0.3578 & 0.5101 & -0.3131 & 0.065 & 0.5371 & 0.4234 & 0.0501 & 0.0 \tabularnewline
			23.043 & 0.0 & -0.0501 & 0.4234 & -0.5371 & 0.065 & 0.3131 & 0.5101 & -0.3578 & -0.1687 & 0.1043 \tabularnewline
	\end{tabular}\end{ruledtabular}
	\label{tableS3}
\end{table*}

\subsection{$\mathrm{La}_{2} \mathrm{PrNi}_{2} \mathrm{O}_{7-\delta}$}

\begin{figure}[htbp]
	\renewcommand\thefigure{S8}
	\includegraphics[width=0.48\textwidth]{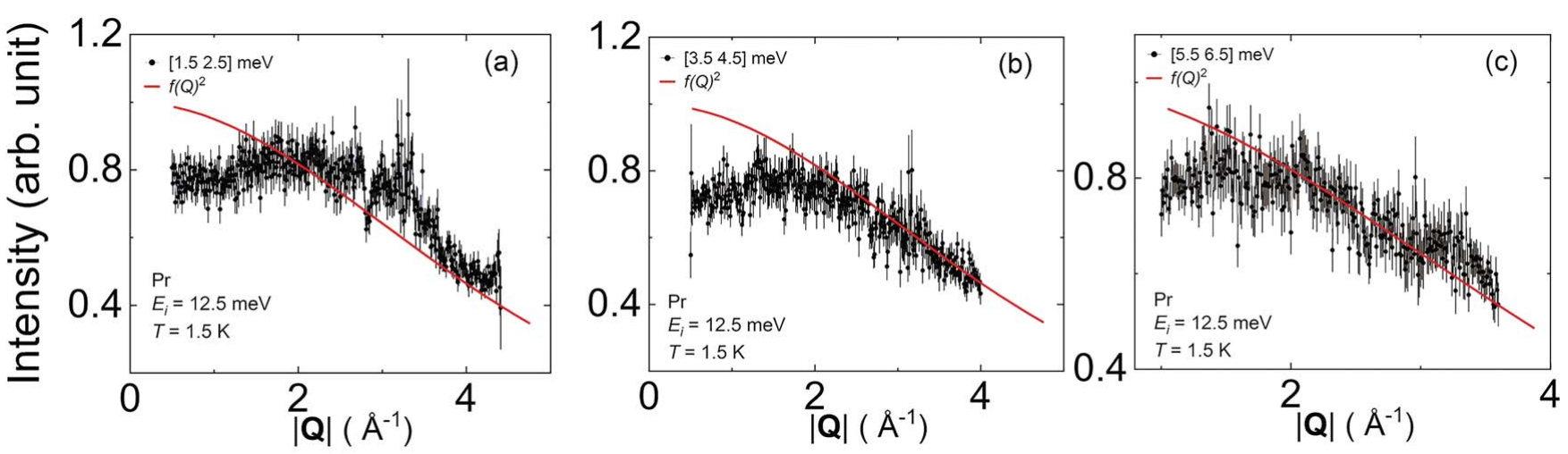}
	\caption{\label{S8}
		$Q$-cuts of the $\mathrm{La}_{2} \mathrm{PrNi}_{2} \mathrm{O}_{7-\delta}$ data with $E_{\mathrm{i}}=12.5$ meV and $T=1.5$ K, integrated over listed energy ranges. The red solid lines are the square of the magnetic form factor for $\mathrm{Pr}^{3+}$ after normalizing by the intensity.
	}
\end{figure}

From the INS results of $\mathrm{La}_{2} \mathrm{PrNi}_{2} \mathrm{O}_{7-\delta}$ measured at $T=$ 1.5 K and 50 K with $E_i=$ 12.5 meV and 19 meV, we can figure out that there are three CEF excitations at 2,4 and 6.5 meV were detected (Fig. S2 and S3). The cuts shown in FIG.\ref{S8}(a)-(c) have $\boldsymbol{I} \propto f(|\boldsymbol{Q}|)^{2}$, where $f(|\boldsymbol{Q}|)$ is the magnetic form factor for $\mathrm{Pr}^{3+}$, and therefore these excitations are mostly identified as CEF levels.

\begin{figure*}[htbp]
	\renewcommand\thefigure{S9}
	\includegraphics[width=0.95\textwidth]{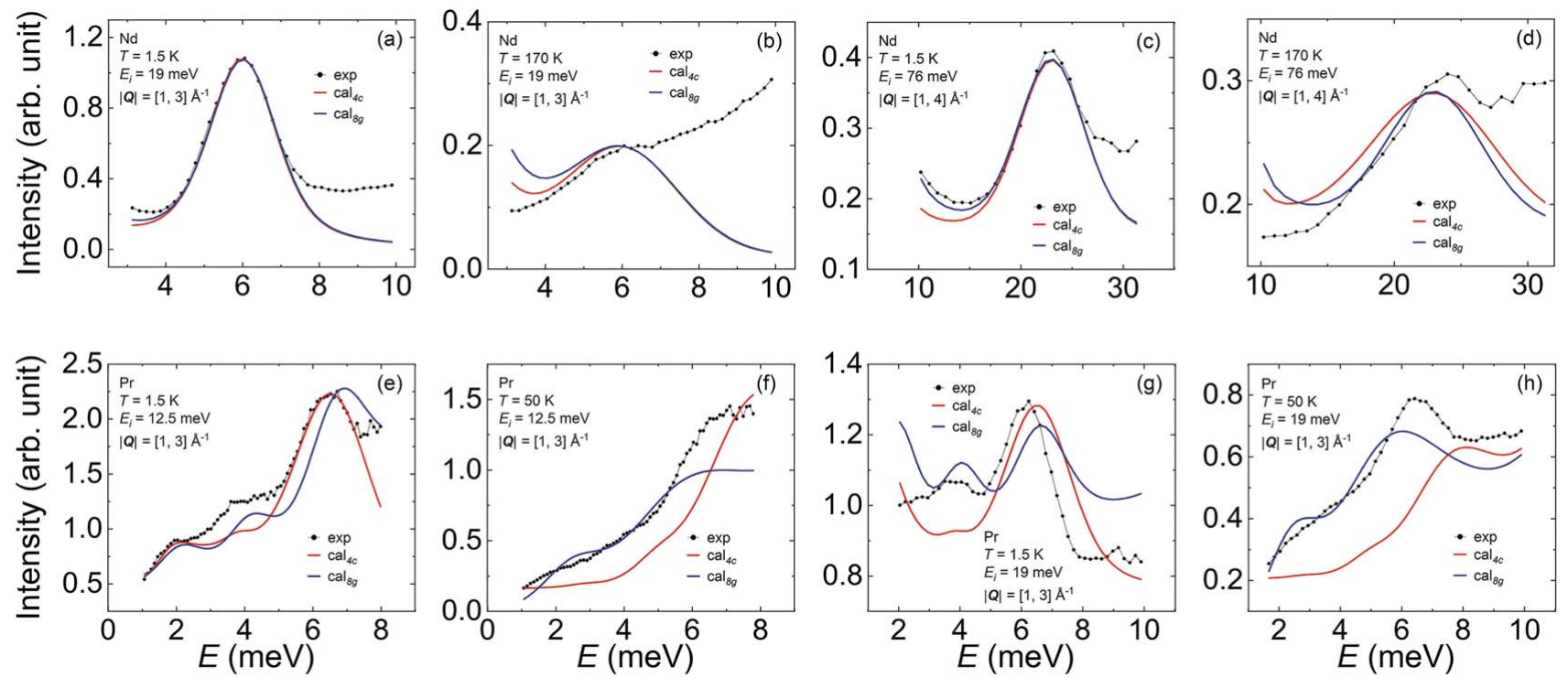}
	\caption{\label{S9}
		Energy cuts of the scattering intensity of $\mathrm{La}_{2} \mathrm{RENi}_{2} \mathrm{O}_{7-\delta}$ obtained from integrating across low $-|\boldsymbol{Q}|$ at different temperatures and $E_{i}$s. The black dots and lines are the raw data from INS experiments. The red and blue solid lines show a fit to the $C_{2 v}$ and $C_{s}$ CEF models as described in the text. 
	}
\end{figure*}

Using Stevens-operator equivalent method, we separately carried out CEF fitting for two different Wyckoff positions of $\mathrm{Pr}^{3+}$ \cite{lyang2025,hnair2017}. The resulting fit curves (red and blue lines) and eigenvalues and eigenvectors are shown in FIG.\ref{S9}(e)-(h), Table \ref{tableS4} and \ref{tableS5}, respectively. The CEF parameters of $\mathrm{La}_{2} \mathrm{PrNi}_{2} \mathrm{O}_{7-8}$ with $\mathrm{Pr}^{3+}$ at $4 c$ Wyckoff positions are $B_{2}^{0}= -5.485 \times 10^{-2} \mathrm{meV}, B_{2}^{2}=-5.633 \times 10^{-2} \mathrm{meV}, B_{4}^{0}=-2.192 \times 10^{-3} \mathrm{meV}, B_{4}^{2}=1.222 \times 10^{-3} \mathrm{meV}, B_{4}^{4}=-5.97 \times 10^{-3} \mathrm{meV}, B_{6}^{0}=-1.583 \times 10^{-5} \mathrm{meV}, B_{6}^{2}=-9.366 \times 10^{-5} \mathrm{meV}$, $B_{6}^{4}=2.273 \times 10^{-4} \mathrm{meV}, B_{6}^{6}=2.128 \times 10^{-5} \mathrm{meV}$. The CEF parameters of $\mathrm{La}_{2} \mathrm{PrNi}_{2} \mathrm{O}_{7-\delta}$ with $\mathrm{Pr}^{3+}$ at $4 c$ Wyckoff positions are $B_{2}^{0}=5.932 \times 10^{-2} \mathrm{meV}, B_{2}^{1}=-1.096 \times 10^{-2} \mathrm{meV}$, $B_{2}^{2}=-2.178 \times 10^{-2} \mathrm{meV}, ~ B_{4}^{0}=7.411 \times 10^{-4} \mathrm{meV}, ~ B_{4}^{1}=3.876 \times 10^{-3} \mathrm{meV}, ~ B_{4}^{2}= -4.562 \times 10^{-4} \mathrm{meV}, B_{4}^{3}=4.092 \times 10^{-3} \mathrm{meV}, B_{4}^{4}=-3.433 \times 10^{-3} \mathrm{meV}, B_{6}^{0}=-6.105 \times 10^{-6} \mathrm{meV}, B_{6}^{1}=-4.421 \times 10^{-5} \mathrm{meV}, B_{6}^{2}=-8.526 \times 10^{-5} \mathrm{meV}, B_{6}^{3}=1.384 \times 10^{-4} \mathrm{meV}$, $B_{6}^{4}=5.693 \times 10^{-4} \mathrm{meV}, B_{6}^{5}=4.846 \times 10^{-4} \mathrm{meV}, B_{6}^{6}=2.321 \times 10^{-4} \mathrm{meV}$.

\begin{table*}
		\renewcommand\thetable{S4}
	\caption{Energy levels and associated wave functions determined from the analysis of the INS data of $\mathrm{La}_{2} \mathrm{PrNi}_{2} \mathrm{O}_{7-\delta}$ with Pr at $4 c$ Wyckoff positions using the $C_{2 v}$ CEF model described in the text.}
	\begin{ruledtabular}
		\begin{tabular}{c|ccccccccc}
			E (meV) &$|-4\rangle$ & $|-3\rangle$ & $|-2\rangle$ & $|-1\rangle$ & $|0\rangle$ & $|1\rangle$ & $|2\rangle$ & $|3\rangle$ & $|4\rangle$ \tabularnewline
			\hline 
			0.000 & 0.7063 & 0.0 & 0.0245 & 0.0 & -0.034 & 0.0 & 0.0245 & 0.0 & 0.7063 \tabularnewline
			0.001 & -0.7067 & 0.0 & -0.0244 & 0.0 & -0.0 & 0.0 & 0.0244 & 0.0 & 0.7067 \tabularnewline
			1.878 & 0.0 & -0.1787 & 0.0 & -0.6842 & 0.0 & -0.6842 & 0.0 & -0.1787 & 0.0 \tabularnewline
			1.936 & 0.0056 & 0.0 & 0.4075 & 0.0 & 0.8172 & 0.0 & 0.4075 & 0.0 & 0.0056 \tabularnewline
			3.806 & 0.0 & 0.2733 & 0.0 & -0.6522 & 0.0 & 0.6522 & 0.0 & -0.2733 & 0.0 \tabularnewline
			3.933 & -0.0339 & 0.0 & 0.5774 & 0.0 & -0.5753 & 0.0 & 0.5774 & 0.0 & -0.0339 \tabularnewline
			6.534 & 0.0 & -0.6842 & 0.0 & 0.1787 & 0.0 & 0.1787 & 0.0 & -0.6842 & 0.0 \tabularnewline
			6.538 & 0.0 & 0.6522 & 0.0 & 0.2733 & 0.0 & -0.2733 & 0.0 & -0.6522 & 0.0 \tabularnewline
			6.568 & -0.0244 & 0.0 & 0.7067 & 0.0 & 0.0 & 0.0 & -0.7067 & 0.0 & 0.0244 \tabularnewline
	\end{tabular}\end{ruledtabular}
	\label{tableS4}
\end{table*}

\begin{table*}
		\renewcommand\thetable{S5}
	\caption{Energy levels and associated wave functions determined from the analysis of the INS data of $\mathrm{La}_{2} \mathrm{PrNi}_{2} \mathrm{O}_{7-\delta}$ with Pr at $8 g$ Wyckoff positions using the $C_{s}$ CEF model described in the text.}
	\begin{ruledtabular}
		\begin{tabular}{c|ccccccccc}
			E (meV) &$|-4\rangle$ & $|-3\rangle$ & $|-2\rangle$ & $|-1\rangle$ & $|0\rangle$ & $|1\rangle$ & $|2\rangle$ & $|3\rangle$ & $|4\rangle$ \tabularnewline
			\hline 
			0.000 & 0.0383 & -0.0475 & -0.6816 & 0.1499 & -0.1354 & -0.1499 & -0.6816 & 0.0475 & 0.0383 \tabularnewline
			1.671 & 0.1218 & 0.3114 & -0.1472 & 0.6054 & 0.0 & 0.6054 & 0.1472 & 0.3114 & -0.1218 \tabularnewline
			1.793 & 0.0285 & -0.5555 & 0.147 & 0.369 & -0.2564 & -0.369 & 0.147 & 0.5555 & 0.0285 \tabularnewline
			2.012 & 0.3376 & 0.2405 & 0.0868 & 0.0233 & -0.8001 & -0.0233 & 0.0868 & -0.2405 & 0.3376 \tabularnewline
			3.792 & 0.0484 & -0.3545 & -0.0788 & -0.5838 & -0.2234 & 0.5838 & -0.0788 & 0.3545 & 0.0484 \tabularnewline
			3.875 & 0.3062 & 0.3323 & 0.5341 & -0.1027 & 0.0 & -0.1027 & -0.5341 & 0.3323 & -0.3062 \tabularnewline
			3.944 & -0.1439 & 0.5308 & -0.3092 & -0.3193 & 0.0 & -0.3193 & 0.3092 & 0.5308 & 0.1439 \tabularnewline
			6.278 & 0.6088 & -0.104 & -0.3122 & -0.1449 & 0.0 & -0.1449 & 0.3122 & -0.104 & -0.6088 \tabularnewline
			6.705 & -0.6176 & 0.0751 & 0.0058 & -0.0067 & -0.4752 & 0.0067 & 0.0058 & -0.0751 & -0.6176 \tabularnewline
	\end{tabular}\end{ruledtabular}
	\label{tableS5}
\end{table*}

From the above analysis on the CEF excitations, we can see the CEF model fitting can capture the peak positions (model energies) but not fully agree with the peak shapes, probably due to some mixed signals from phonons or magnons. According to the spin-wave calculation results in Fig.1, there are still some magnon contributions at low energy and low $|\boldsymbol{Q}|$ range, but probably their spectrum weight is too weak to detect by powder INS. We have tried to use the above CEF parameters to fit the 45 meV and 60 meV models, but not successfully. As previous INS on undoped $\mathrm{La}_{3} \mathrm{Ni}_{2} \mathrm{O}_{7-\delta}$ has already shown the spin excitations emerged at 45 meV, and two splitting branches in Nd doped sample actually show different $|\boldsymbol{Q}|$ dependence with increasing splitting energies, we have a good reason to believe they are magnon excitations. Further, from the discussion with Jun Zhao's group at Fudan University, we compare with their single crystal INS results, and conclude that the splitting modes are actually the different band tops in spin wave branches, which has been illustrated in the previous section of spin-wave calculations (Fig. S6).

\end{document}